\documentclass[acmtog,nonacm,screen,natbib=false]{acmart}
\input{preamble}
\acmArticle{}
\begin{document}

\title{Gaussian Process Implicit Surfaces as Participating Media: Realization-Free Rendering from Level-Crossing Statistics}

\author{Jack Cui}
\affiliation{%
  \institution{Dartmouth College}
  \city{Hanover}
  \country{USA}
}
\orcid{0009-0008-4763-2861}
\email{jack.cui.gr@dartmouth.edu}

\author{Kehan Xu}
\affiliation{
    \institution{Dartmouth College}
    \country{USA}
}
\orcid{0009-0009-7860-3593}
\email{kehan.xu.gr@dartmouth.edu}

\author{Eugene d'Eon}
\affiliation{
    \institution{NVIDIA}
    \country{Switzerland}
}
\orcid{0000-0002-3761-2989}
\email{edeon@nvidia.com}

\author{Wojciech Jarosz}
\affiliation{%
  \institution{Dartmouth College}
  \city{Hanover}
  \country{USA}
}
\orcid{0000-0002-1652-0954}
\email{wojciech.k.jarosz@dartmouth.edu}

\renewcommand{\shortauthors}{Jack Cui, Kehan Xu, Eugene d'Eon, and Wojciech Jarosz}

\begin{abstract}

We present a theory of light scattering that connects Gaussian Process
Implicit Surfaces (GPISes) and participating media in both directions.
Applying the Kac--Rice level-crossing formula under a local-conditioning
approximation yields a complete anisotropic radiative transfer equation
(RTE) directly from pointwise GPIS statistics.
A shared projected area couples extinction and scattering,
ensuring geometric consistency between the GPIS and its volumetric
representation.
The framework spans rough surfaces, porous and non-height-field
geometries, and participating media.
From the same statistical structure, we derive full-sphere Beckmann
and GGX normal distribution functions supporting in-plane and
out-of-plane anisotropy.
These families provably recover SGGX, Beckmann, and GGX as special
cases and admit exact visible-normal importance sampling.
We also derive analytic masking--shadowing functions and
single-scattering surface models for specular microsurfaces,
with extensions to multiple scattering.
In the height-field limit, we prove that the local-conditioning approximation reduces
to Smith's independence assumption.
Our realization-free approach improves rendering efficiency over
realization-based methods and can be implemented within a
standard volume renderer.
In the inverse direction, we characterize families of GPISes
corresponding to compatible RTE parameters and develop practical lifts
for heterogeneous density fields.
Existing volumetric assets thereby become renderable as GPISes, while
trained radiance-field reconstructions yield surface geometry and
shading normals without mesh extraction and provide a density-based
representation of geometric uncertainty.

\end{abstract}



\begin{teaserfigure}
  \includegraphics[width=\textwidth]{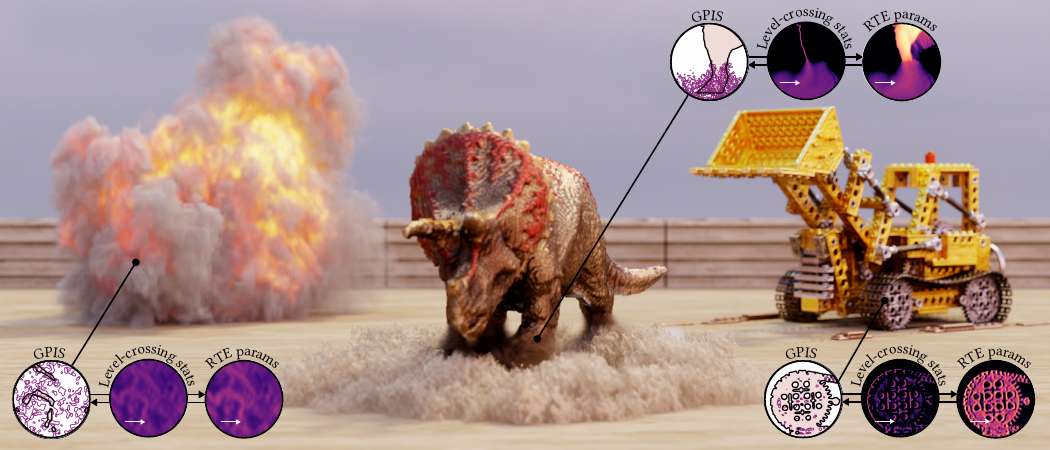}
  \caption{From surfaces to media---and back. Using Kac--Rice
level-crossing theory with a local-conditioning approximation, we
derive anisotropic RTE parameters directly from pointwise GPIS
statistics, enabling rendering without sampling explicit geometry
realizations. This formulation spans participating media (left),
porous-to-surface transitions (center), and hard surfaces (right); the
insets follow each GPIS through its level-crossing statistics to the
resulting RTE parameters (and back). Conversely, we characterize GPIS families
corresponding to prescribed RTE parameters, lifting classical volumes
to probabilistic surfaces.}
  \Description{teaser}
  \label{fig:teaser}
\end{teaserfigure}

\maketitle

\section{Introduction}
\label{sec:introduction}

Gaussian Process Implicit Surfaces (GPISes) are a natural probabilistic representation
for 3D geometry with uncertainty: each realization of a Gaussian process defines a
possible surface through its zero-level set.
Despite this appeal, rendering GPISes is expensive.
State-of-the-art methods sample and ray trace explicit geometry
realizations~\citep{Seyb:2024:Microfacets,Xu:2025:Practical}, paying for
both realization generation and intersection.
Classical volumetric representations, governed by the radiative transfer equation
(RTE)~\citep{Chandrasekhar:1960:Radiative, Pharr:2023:Physically}, offer the opposite
trade-off: fast, differentiable rendering with a rich algorithmic ecosystem, but no
explicit probabilistic surface model.
The central question of this paper is whether---and how precisely---these two
representations can be converted into one another.

We derive a bidirectional conversion between GPISes and participating
media (\cref{fig:teaser}).
The forward direction (GPIS $\to$ RTE, \cref{sec:forward})
replaces realization-based rendering with closed-form extinction
evaluation and importance-sampled scattering.
This allows GPISes to benefit from established volume-rendering
techniques, including next-event estimation and differentiable rendering.
The inverse direction (RTE $\to$ GPIS, \cref{sec:inversion})
characterizes families of GPISes corresponding to compatible target
RTE parameters.
This gives volumetric representations a probabilistic surface
interpretation.

Exact GPIS transport depends on spatial correlations along the ray's
traversal history, whereas a classical RTE is memoryless and governed
by local coefficients.
We therefore adopt a \emph{local-conditioning approximation}:
rather than conditioning a surface encounter on the ray having remained
in empty space up to the current point, we condition only on the current
point being in empty space.
The resulting transport model depends only on pointwise field and
gradient statistics.
The approximation preserves the exact initial crossing rate for rays
conditioned only to start in empty space.
Under stationarity, the approximation also preserves the mean
chord length in the void.
We quantify the approximation error against explicit GPIS realizations
and show that, in the height-field limit, it reduces to Smith's
independence assumption~\citep{Smith:1967:Geometrical}, recovering the
corresponding shadowing function exactly.

Our derivation is \emph{3D-first}: we begin with a general nonstationary
GPIS and apply the Kac--Rice level-crossing
formula~\citep{rice1945mathematical,Kac:1943:Average} to its pointwise
field and gradient statistics.
This yields closed-form extinction and full-sphere normal distribution
functions (NDFs), coupled through a common projected area that ensures
consistency between attenuation and surface-normal
statistics.
The full-sphere support accommodates normals arising from overhangs
and porous geometry beyond height fields.
The centered case recovers SGGX exactly~\citep{Heitz:2015:SGGX},
while the height-field limit recovers the Beckmann NDF.

Our contributions, in the order developed in the paper, are:
\begin{itemize}
    \item \textbf{Forward conversion.}
        An anisotropic RTE for general GPISes under local conditioning,
        with closed-form extinction
        (\cref{sec:forward}).

    \item \textbf{Normal distributions and surface scattering.}
        Full-sphere Beckmann and GGX NDF families supporting porous media, with analytic masking--shadowing
        functions and single-scattering surface models for specular
        microsurfaces, and extensions to multiple scattering
        (\cref{sec:ndfderivation,sec:mixture,sec:surfaceconnections}).

    \item \textbf{Sampling and rendering.}
        Exact gradient-space vNDF sampling for both families and a
        realization-free renderer achieving $23$--$34\times$ lower
        equal-time estimator MSE than the realization-based
        baseline~\citep{Xu:2025:Practical}
        (\cref{sec:rendering,sec:results}).

    \item \textbf{Approximation analysis.}
        A proof that local conditioning reduces to Smith's assumption
        in the height-field limit, a qualitative correlation-to-chord
        accuracy diagnostic, and validation against explicit GPIS
        realizations
        (\cref{sec:lca_accuracy,app:beckmannsmith}).

    \item \textbf{Inverse conversion.}
        GPIS equivalence families for compatible RTE targets, with
        practical lifts of volumetric assets and radiance-field
        reconstructions
        (\cref{sec:inversion,sec:results}).
\end{itemize}

\section{Related work}
\label{sec:relatedwork}

\paragraph{Rendering with stochastic geometry}
\citet{Seyb:2024:Microfacets} introduced ensemble-averaged light
transport for GPISes and a renderer based on sampling realizations
along ray segments.
Their formulation connects microfacet surfaces and participating
media through a common stochastic geometry representation.
\citet{Xu:2025:Practical} improved rendering efficiency
using sparse-convolution noise and next-event estimation, while
subsequent work further accelerates realization-based ray
marching~\citep{Shi:2026:Conditional,Chen:2026:Bernoulli,Zhou:2026:Adaptive}.
These methods retain realization sampling and intersection,
whereas our local-conditioning approximation enables
realization-free rendering from pointwise GPIS statistics.

\citet{Miller:2024:Objects} established conditions under which
stochastic opaque solids admit exponential volumetric transport,
while noting that GPIS free-flight distributions are generally
non-exponential.
Our local-conditioning approximation discards ray history,
yielding an exponential transport model governed by a classical RTE.

\paragraph{Microfacet and microflake theory}
Microfacet models~\citep{Beckmann:1963:Scattering,Cook:1981:Reflectance,
Walter:2007:Microfacet} represent rough surfaces through an 
NDF and a masking--shadowing function.
\citet{Seyb:2024:Microfacets} connected height-field GPISes to
Beckmann microfacets and demonstrated agreement empirically.
We show analytically that, in the height-field limit, our full-sphere
NDF reduces to Beckmann and local conditioning to Smith's
independence assumption~\citep{Smith:1967:Geometrical}
(\cref{app:beckmannlimit,app:beckmannsmith}).

\citet{Heitz:2014:Understanding} analyzed the masking-shadowing
function and derived the projected-area normalization of the visible NDF.
\citet{Seyb:2024:Microfacets} obtained a vNDF for general GPISes as an
integral of their crossing density. Under local conditioning, we recover
the same projected-area form in closed form, with the NDF and its
normalization arising as moments of a single conditional gradient law.

\citet{Jakob:2010:Radiative} extended the RTE to anisotropic microflake
media, providing the framework our rendering equation directly instantiates.
\citet{Heitz:2015:SGGX} proposed the SGGX microflake distribution. We
show that a zero conditional mean gradient
produces SGGX exactly and that the general GPIS NDF converges to it in
the high-roughness limit, thereby giving SGGX a probabilistic derivation
from GPIS statistics (\cref{sec:ndfproperties}).

\citet{Heitz:2016:Multiplescattering} extended the Smith microfacet
model to multiple scattering through random walks in an inhomogeneous
microflake medium.
\citet{Dupuy:2016:Additional} remapped this construction to an
equivalent homogeneous half-space.
We extend both formulations to BRDF-compatible
GPISes with non-height-field geometry where
one-sided interactions arise naturally from oriented GPIS boundary
crossings (\cref{sec:multiple-scattering-walk}).

Full-sphere NDFs were proposed to connect rough surfaces with
semi-porous media beyond height-field geometry~\citep{Dupuy:2016:Additional}.
Subsequent work explored von Mises--Fisher
NDFs~\citep{dEon:2016:Anisotropic,dEon:2024:VMF},
whose projected areas lack a known simple closed form,
and developed null-scattering methods for general full-sphere
NDFs~\citep{dEon:2023:StudentT}.
Our full-sphere Beckmann and GGX families provide closed-form
one-sided projected areas and exact vNDF sampling
(\cref{sec:fullsphere-ndfs}).

\paragraph{Random surfaces and stochastic media}
Kac--Rice theory describes level crossings of random
fields~\citep{Kac:1943:Average,rice1945mathematical,Azais:2009:Level}.
Related work connects chord-length distributions in random media
to first-passage times of Gaussian
processes~\citep{Roberts:1999:Chorddistribution},
while higher-order Rice expansions retain crossing
history~\citep{Lindgren:2019:Gaussian,Kapp:1994:Effect}.
The same local-conditioning approximation we use has been applied
to Gaussian height fields~\citep{Lumme:1990:Diffuse} and level sets
of 3D Gaussian fields~\citep{Peltoniemi:1993:Radiative},
with spatially uniform roughness statistics.
We extend this construction to general GPISes
(\cref{sec:forward}).

\paragraph{Learned scene representations}
NeRF~\citep{Mildenhall:2020:NeRF} learns volumetric density and
view-dependent radiance from images, while
3D Gaussian splatting~\citep{Kerbl:2023:3D} learns Gaussian
primitives with opacity and appearance.
Our inverse conversion allows learned density fields to be reused
as probabilistic surface geometry.
We demonstrate this on TensoRF~\citep{Chen:2022:TensoRF}
reconstructions, converting their density grids into renderable
GPISes with surface geometry and shading normals, without mesh
extraction or another reconstruction stage.
The resulting representation also exposes geometric ambiguity
encoded in the learned density.

\paragraph{Concurrent work}
\citet{Huang:2026:Macrofacet} also derive a realization-free
GPIS rendering model by extending microfacet theory into a volume
around a macro-surface.
For their linear-ramp mean and stationary, axis-aligned
squared-exponential kernel, the extinction coefficient, NDF,
and vNDF coincide algebraically with ours.
Our derivation handles general nonstationary GPISes and additionally
provides exact vNDF sampling and an inverse conversion from
RTE parameters to GPISes.
We also derive analytic masking--shadowing and single-scattering
surface models for specular microsurfaces with full-sphere
Beckmann and GGX NDFs, and extend these models to multiple scattering
(\cref{sec:discussion}).
\citet{Genest:2026:Uncertainty} independently apply the Kac--Rice
formula to GPISes for uncertainty-aware geometry processing.

\begin{table}[t]
  \caption{Commonly used notation.}
  \label{tab:notation}
  \centering
  \small
  \begin{tabular}{@{}ll@{}}
    \toprule
    $\mathbf{a},\;\hat{\mathbf{a}},\;\mathbf{A}$
      & Vector; unit vector; $3\times3$ matrix \\
    $\langle\mathbf{a},\mathbf{b}\rangle$
      & Clamped dot product \\
    $\normPDF,\;\normCDF$
      & Standard normal PDF and CDF \\
    \midrule
    $\RayDir,\;\ScatDir$
      & Ray directions before and after scattering \\
    $\OutDir[i],\;\OutDir[o]$
      & Outward incident and outgoing BRDF directions \\
    $\SurfNorm,\;\HalfVec$
      & Surface normal; half-vector \\
    \midrule
    $\GPField$
      & GPIS field \\
    $\GPMean,\;\StdF$
      & Field mean and standard deviation \\
    $\CovFn$
      & Field covariance kernel \\
    $\VoidLevel,\;\VoidFrac$
      & Void level; void fraction \\
    $\GradField,\;\CondGrad$
      & Gradient; zero-conditioned gradient \\
    $\MeanGrad,\;\CovGrad$
      & Gradient mean and covariance \\
    $\CrossCov$
      & Field--gradient cross-covariance \\
    $\CondGradMean,\;\CondGradCov$
      & Conditional gradient mean and covariance \\
    \midrule
    $\DirDeriv,\;\CondDeriv$
      & Ray derivative; zero-conditioned derivative \\
    $\CondDerivMean,\;\CondDerivStd$
      & Mean and standard deviation of $\CondDeriv$ \\
    \midrule
    $\DownCross$
      & Down-crossings per unit ray length \\
    $\ProjAreaMinus,\;\ProjAreaPlus$
      & Down-/up-crossing projected areas \\
    $\ExtCoeff,\;\ExtCoeffExact$
      & Local and exact extinction \\
    $\Trans,\;\TransExact$
      & Local and exact transmittance \\
    $\FreeFlightPDF$
      & Free-flight PDF \\
    $\Radiance,\;\RadianceBG$
      & Radiance; background radiance \\
    \midrule
    $\NDF,\;\VNDF[\RayDir]$
      & NDF; visible NDF \\
    $\PhaseFunc,\;f_r$
      & Phase function; micro-BRDF \\
    $\alpha$
      & Roughness parameter \\
    $\SmithLambda,\;\ShadowG,\;G_{12}$
      & Smith auxiliary; masking--shadowing functions \\
    \midrule
    $\CorLen$
      & Kernel correlation length \\
    $C_{\mathrm v},\;\eta$
      & Void-chord length; correlation-to-chord ratio \\
    $\SGGXMat,\;\SGGXDensity$
      & SGGX shape matrix; medium density \\
    \bottomrule
  \end{tabular}
\end{table}

\section{Preliminaries and Notation}
\label{sec:preliminaries}

We aim to estimate the \emph{ensemble-averaged radiance}, the expectation of
radiance over realizations of the GPIS \citep{Seyb:2024:Microfacets}.
Rather than sampling those realizations explicitly
\citep{Seyb:2024:Microfacets,Xu:2025:Practical}, we approximate their
ensemble-averaged transport using a single anisotropic participating medium.
This medium is specified by an extinction coefficient $\ExtCoeff$ and a phase
function $\PhaseFunc$, both derived from the pointwise GPIS statistics developed
below.

Throughout the paper, we use bold lowercase letters to denote 3D vectors and bold uppercase letters for
$3\times3$ matrices; plain lowercase letters denote scalars. A hat denotes a unit
vector, $\hat{\mathbf{a}} \coloneq \mathbf{a}/\|\mathbf{a}\|$. We write
$\langle\mathbf{a},\mathbf{b}\rangle \coloneq \max(\mathbf{a}\cdot\mathbf{b},\,0)$
for the dot product clamped to zero. \Cref{tab:notation} summarizes commonly used
notation.

\subsection{The 3D Field and Gradient Statistics}
\label{sec:fieldgradientstats}

A GPIS is a continuous, generally non-stationary Gaussian field
$\GPField[\mathbf{x}]$, $\mathbf{x} \in \mathbb{R}^3$, fully characterized by its
mean function $\GPMean[\mathbf{x}] = \mathbb{E}[\GPField[\mathbf{x}]]$ and
covariance function
$\CovFn[\mathbf{x}][\mathbf{x}'] = \mathrm{Cov}(\GPField[\mathbf{x}], \GPField[\mathbf{x}'])$;
by convention $\GPField > 0$ indicates empty space, $\GPField < 0$ solid mass, and
$\GPField = 0$ the implicit surface boundary. By linearity of the derivative
operator, the spatial gradient
\begin{math}
  \GradField[\mathbf{x}] \coloneq \nabla\GPField[\mathbf{x}]
\end{math}
is itself a Gaussian process, jointly Gaussian with $\GPField[\mathbf{x}]$ at every
point, and the surface normal is the normalized gradient,
$\SurfNorm = \GradField[\mathbf{x}]/\|\GradField[\mathbf{x}]\|$, which points toward increasing $\GPField$ and hence
into empty space.

We write the $d$-dimensional Gaussian PDF as
\begin{align}
  \label{eqn:gaussian_pdf}
  \mathcal{N}(\mathbf{x};\,\boldsymbol{\mu},\,\boldsymbol\Sigma)
    \coloneq
    (2\pi)^{-d/2}|\boldsymbol\Sigma|^{-1/2}
    \exp\!\left(-\tfrac{1}{2}
      (\mathbf{x}-\boldsymbol{\mu})^\mathsf{T}\boldsymbol\Sigma^{-1}(\mathbf{x}-\boldsymbol{\mu})
    \right),
\end{align}
and define $\normPDF[x] \coloneq \mathcal{N}(x;\,0,\,1) = \frac{1}{\sqrt{2\pi}}\euler^{-x^2/2}$
and $\normCDF[x] \coloneq \int_{-\infty}^{x}\normPDF[s]\,\dif s$ as the PDF and CDF of the
standard normal.

The joint distribution of $\GPField[\mathbf{x}]$ and $\GradField[\mathbf{x}]$ at any
point $\mathbf{x}$ is the 4D Gaussian \citep[Sec.~9.4, Eq.~(9.1)]{Rasmussen:2006:Gaussian}:
\begin{align}
  \label{eqn:joint_field_gradient}
  \begin{bmatrix} \GPField[\mathbf{x}] \\ \GradField[\mathbf{x}] \end{bmatrix}
  \sim \mathcal{N}\!\left(
    \begin{bmatrix} \GPMean[\mathbf{x}] \\ \MeanGrad[\mathbf{x}] \end{bmatrix},\;
    \begin{bmatrix}
      \FieldVar[\mathbf{x}] & \CrossCov[\mathbf{x}][\mathsf{T}] \\
      \CrossCov[\mathbf{x}] & \CovGrad[\mathbf{x}]
    \end{bmatrix}
  \right),
\end{align}
with parameters derived from the mean and covariance functions:
\begin{itemize}
  \item $\MeanGrad[\mathbf{x}] \coloneq \nabla\GPMean[\mathbf{x}]$: mean of the gradient
        process $\GradField[\mathbf{x}]$, equal to the gradient of the field mean
        by linearity of expectation.
  \item $\FieldVar[\mathbf{x}] = \CovFn[\mathbf{x}][\mathbf{x}]$: variance of the
        field value.
  \item $\CovGrad[\mathbf{x}] =
        \nabla_{\mathbf{x}}\nabla_{\mathbf{x}'}^{\mathsf{T}}\CovFn[\mathbf{x}][\mathbf{x}']\big|_{\mathbf{x}'=\mathbf{x}}$:
        $3\times 3$ covariance matrix of $\GradField[\mathbf{x}]$. For any stationary kernel $\CovFn[\mathbf{x}][\mathbf{x}'] = k(\mathbf{x}-\mathbf{x}')$ this reduces to $\CovGrad = -(\nabla\nabla^{\mathsf{T}} k)(\mathbf{0})$.
  \item $\CrossCov[\mathbf{x}] =
        \nabla_{\mathbf{x}'}\CovFn[\mathbf{x}][\mathbf{x}']\big|_{\mathbf{x}'=\mathbf{x}}$:
        cross-covariance vector between $\GPField[\mathbf{x}]$ and $\GradField[\mathbf{x}]$.
\end{itemize}

Surface intersections and scattering events occur on the zero set
$\GPField[\mathbf{x}] = 0$. We write
$\CondGrad[\mathbf{x}] \coloneq \GradField[\mathbf{x}] \mid \GPField[\mathbf{x}] = 0$
for the gradient conditioned on a zero field value; the bar marks conditioning
throughout, and statistics of $\CondGrad$ carry the subscript $\bar{\mathbf{g}}$, as
in $\CondGradMean$ and $\CondGradCov$. This conditions on the field value only---the
distribution of gradients \emph{observed at a crossing} carries an additional slope
weight, which we consider in \cref{sec:ndfderivation}. By the standard Gaussian conditioning
formula, $\CondGrad[\mathbf{x}]$ is a 3D Gaussian with:
\begin{subequations}
\label{eqn:grad_cond_stats}
\begin{align}
  \label{eqn:grad_mean_cond}
  \CondGradMean[\mathbf{x}]
    &\coloneq
    \MeanGrad[\mathbf{x}]
       - \frac{\CrossCov[\mathbf{x}]}{\FieldVar[\mathbf{x}]}
         \GPMean[\mathbf{x}],\\
  \label{eqn:grad_cov_cond}
  \CondGradCov[\mathbf{x}]
    &\coloneq \CovGrad[\mathbf{x}]
       - \frac{\CrossCov[\mathbf{x}]\,\CrossCov[\mathbf{x}][\mathsf{T}]}{\FieldVar[\mathbf{x}]}.
\end{align}
\end{subequations}
Note that the cross-covariance is not an independent quantity: differentiating
$\FieldVar[\mathbf{x}] = \CovFn[\mathbf{x}][\mathbf{x}]$ and using the
symmetry of $\CovFn$ gives
\begin{align}
  \label{eqn:crosscov_identity}
  \CrossCov[\mathbf{x}]
    = \tfrac{1}{2}\nabla\FieldVar[\mathbf{x}]
    = \StdF[\mathbf{x}]\,\nabla\StdF[\mathbf{x}].
\end{align}
Substituting \cref{eqn:crosscov_identity} into
\cref{eqn:grad_cond_stats} gives the conditional statistics directly
in terms of the mean and variance fields:
\begin{align}
  \label{eqn:condgrad_identities}
  \CondGradMean
    = \StdF\,\nabla\!\left(\frac{\GPMean}{\StdF}\right),
  \qquad
  \CondGradCov
    = \CovGrad - \nabla\StdF\,\nabla\StdF^{\mathsf{T}}.
\end{align}
For stationary kernels, the variance is constant, so $\CrossCov =
\mathbf{0}$ and conditioning changes nothing: $\CondGradMean =
\MeanGrad$ and $\CondGradCov = \CovGrad$.

\subsection{Ray Restriction}
\label{sec:rayrestriction}

We denote a ray with origin $\mathbf{o}$ and unit direction $\RayDir$ as
$\mathbf{r}(t) = \mathbf{o} + t\RayDir$.
Restricting the GPIS to it gives the 1D
field process $\GPField[t] \coloneq \GPField[\mathbf{r}(t)]$ \citep{Seyb:2024:Microfacets,Roberts:1999:Chorddistribution}.
The scalar directional derivative along the ray,
\begin{align}
  \label{eqn:dir_deriv_def}
  \DirDeriv[t]
    \coloneq \RayDir^\mathsf{T}\GradField[\mathbf{r}(t)]
    = \frac{\partial}{\partial t}\GPField[t],
\end{align}
is jointly Gaussian with $\GPField[t]$.
The pair $(\GPField[t], \DirDeriv[t])$ is characterized by five scalar statistics,
obtained from the corresponding 3D quantities at $\mathbf{r}(t)$, the vector ones
projected onto $\RayDir$:
\begin{itemize}
  \item $\MeanF[t] = \GPMean[\mathbf{r}(t)]$: mean field value along the ray.
  \item $\MeanFp[t] = \RayDir^\mathsf{T}\MeanGrad[\mathbf{r}(t)]$: mean of $\DirDeriv[t]$.
  \item $\VarF[t] = \CovFnRay[t][t] \coloneq \CovFn[\mathbf{r}(t)][\mathbf{r}(t)]$: variance of $\GPField[t]$.
  \item $\VarFp[t]
        = \!\left.\dfrac{\partial^2 \CovFnRay[t][t']}{\partial t\,\partial t'}\right|_{t'=t}
        \!\!= \RayDir^\mathsf{T}\CovGrad[\mathbf{r}(t)]\,\RayDir$:
        variance of $\DirDeriv[t]$.
  \item $\CovFFp[t]
        = \!\left.\dfrac{\partial \CovFnRay[t][t']}{\partial t'}\right|_{t'=t}
        \!\!= \RayDir^\mathsf{T}\CrossCov[\mathbf{r}(t)]$:
        covariance between $\GPField[t]$ and $\DirDeriv[t]$.
\end{itemize}

Similarly, we write
$\CondDeriv[t] \coloneq \DirDeriv[t] \mid \GPField[t] = 0$ for the directional
derivative conditioned on a zero field value, a 1D Gaussian with mean
$\CondDerivMean[t]$ and variance $\CondDerivVar[t]$ given by projecting
\cref{eqn:grad_cond_stats}:
\begin{subequations}
\label{eqn:ray_cond_stats}
\begin{align}
  \label{eqn:ray_cond_mean}
  \CondDerivMean[t]
    &= \RayDir^\mathsf{T} \CondGradMean[\mathbf{r}(t)]
     = \MeanFp[t] - \frac{\CovFFp[t]}{\VarF[t]}\MeanF[t],\\
  \label{eqn:ray_cond_cov}
  \CondDerivVar[t]
    &= \RayDir^\mathsf{T} \CondGradCov[\mathbf{r}(t)]\, \RayDir
     = \VarFp[t] - \frac{\CovFFp[t][2]}{\VarF[t]}.
\end{align}
\end{subequations}

A ray traveling through empty space meets a GPIS realization at a
\emph{down-crossing}: a point where $\GPField[t] = 0$ with $\DirDeriv[t] < 0$,
equivalently where the normal faces the ray, $\SurfNorm\cdot\RayDir < 0$. A minus subscript marks quantities tied to
down-crossings and a plus subscript those tied to up-crossings. 
We assume GPIS realizations are smooth enough that surface normals
are well defined and ray crossings are isolated.\footnote{Formally, we assume the field
has almost-surely continuously differentiable sample paths; the zeros
of each ray-restricted process are simple
($\GPField[t]=0$ implies $\DirDeriv[t]\neq0$); and the relevant joint
field--derivative Gaussian distributions have nonsingular covariance.
These are standard sufficient conditions for the Kac--Rice formulas
used below \citep{Azais:2009:Level}. Without them, normals or crossing
rates may be undefined and the crossing formulas need not apply.}

\subsection{Anisotropic RTE}
\label{sec:rte}
In an anisotropic participating medium \citep{Jakob:2010:Radiative}, the radiance arriving at a point $\mathbf{o}$ along direction $-\RayDir$ is:
\begin{align}
  \label{eq:re}
  \!\!\Radiance[\mathbf{o}][\RayDir]
    =&{} \int_0^\infty\!\!\!
        \underbrace{\ExtCoeff[t]\,\Trans[t]}_{\FreeFlightPDF[t]}
        \left[\int_{S^2}
          \PhaseFunc[\RayDir][\ScatDir]\,
          \Radiance[\mathbf{r}(t)][\ScatDir]\,
          \dif\ScatDir
        \right] \dif t \!\!\notag\\
      +&{}\, \Trans[\infty]\, \RadianceBG[\RayDir].
\end{align}
Here, $\Trans[t]$ is the transmittance, $\ExtCoeff[t]$ is the extinction
coefficient, and their product is the free-flight PDF $\FreeFlightPDF[t]$. The direction after scattering is $\ScatDir$
 and $\RadianceBG[\RayDir]$ is the background
radiance.
Each crossing is governed by a prescribed micro-BRDF, assumed to be deterministic
(independent of the field fluctuations) though it may vary with position. If it
absorbs, the phase function is subnormalized, integrating to the single-scattering
albedo, $\int_{S^2} \PhaseFunc[\RayDir][\ScatDir]\,\dif\ScatDir \le 1$.

The RTE \eqref{eq:re} is memoryless: radiance evolves through local coefficients
alone. A GPIS is not---field values the ray has already observed are correlated with
the geometry ahead. Any RTE representation therefore discards some of this information,
and many distinct GPISes can induce the same local transport coefficients.
We estimate them using a \emph{local-conditioning approximation}
(\cref{sec:approx}).

\section{From GPISes to Participating Media}
\label{sec:forward}

We now show how to represent a GPIS as an anisotropic participating medium. To
do this, we need to know where a ray next
encounters the surface and how it scatters there. Kac--Rice theory lets us answer
both questions from the same pointwise field and gradient statistics. We begin
with the exact probability of a surface encounter and then introduce the local
approximation needed to obtain tractable transport coefficients.

\subsection{Exact Transmittance and Extinction}
\label{sec:ode}

Let $\TransExact[t]$ denote the exact ray transmittance: the probability that a ray beginning at $\mathbf{o}$ in empty space and traveling in direction $\RayDir$ remains in empty space up to depth $t$:
\begin{align}
  \TransExact[t]
  = \mathbb{P}\!\left(
      \GPField[s] > 0 \;\;\forall\, s \in (0,t]
      \;\middle|\;
      \GPField[0] > 0
    \right),
\end{align}
so that $\TransExact[0]=1$. Equivalently, $\TransExact[t]$ is called the survival function 
of the distance to the first surface intersection~\citep{durbin1985first}.

\begin{figure*}[t]
  \centering
  \includegraphics[width=\textwidth]{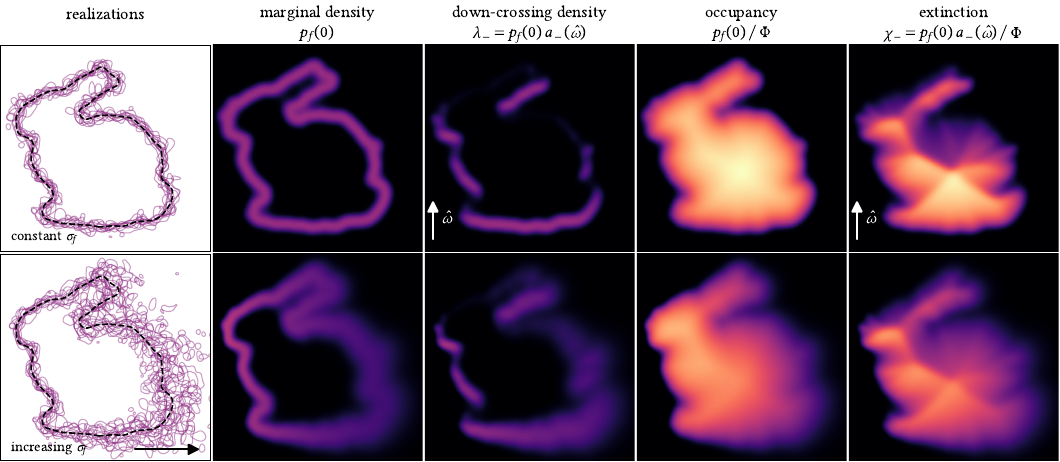}
  \caption{From a distribution over surfaces to closed-form transport
    coefficients. The mean surface is the bunny outline; $\StdF$ is constant in
    the top row and increases left to right in the bottom row. Realizations
    (purple) wiggle around the mean surface (dashed). The marginal density of a
    zero field value, $p_{\GPField}(0)$ \eqref{eqn:downcrossing_factored}, widens
    where the surface is uncertain. The down-crossing density $\DownCross$
    \eqref{eqn:kacrice_density} weights it by the one-sided projected area along
    $\RayDir$: only ray-facing surfaces contribute. Dividing each by
    $\normCDF[\MeanF/\StdF]$, the probability that the point lies in empty space,
    gives the occupancy factor of \cref{eqn:extinction_coeff} and the extinction
    coefficient $\ExtCoeff$, which increases inside the solid. All density maps 
    share one asinh color scale, linear through the surface band and compressive beyond.}
  \label{fig:kac_rice_bunny}
\end{figure*}

The corresponding extinction rate $\ExtCoeffExact[t]$ is the instantaneous rate of a first 
down-crossing, conditioned on the ray having remained in empty space up to depth $t$. 
Consequently, $\TransExact[t]$ satisfies the ODE:
\begin{align}
  \label{eqn:transmittance_ode}
  \frac{\dif\TransExact[t]}{\dif t} &= -\ExtCoeffExact[t]\,\TransExact[t],\quad\text{where}\\
  \label{eqn:non_local_extinction}
  \ExtCoeffExact[t]
    &= \lim_{\Delta t \to 0}
      \tfrac{\mathbb{P}\left(\text{down-crossing in }[t,\,t+\Delta t]
            \;\mid\; \GPField[s] > 0 \;\forall\, s \in [0, \,t]\right)}{\Delta t}.
\end{align}
A down-crossing is a transition from empty space to solid mass:
$\GPField[t] = 0$ with $\DirDeriv[t] < 0$.
Note that as defined above, $\ExtCoeffExact[t]$ is not just a function of the point at $t$,
but of the entire history along the ray from $0$ to $t$.  Restricted
to a height field, this setup reduces to the exact ODE from which
classical rough-surface shadowing derivations begin
\citep{Beckmann:1965:Shadowing,Wagner:1967:Shadowing,Smith:1967:Geometrical}.

\subsection{The Local-Conditioning Approximation}
\label{sec:approx}

Evaluating the exact extinction rate requires conditioning on the full path history which is analytically intractable in general.
To obtain a local coefficient, we discard this history and 
condition only on the current point being in empty space. We call this the 
\emph{local-conditioning approximation} (LCA):
\begin{align}
  \ExtCoeffExact[t]
    &\approx \ExtCoeff[\mathbf{r}(t),\RayDir] \notag \\ 
    &= \lim_{\Delta t \to 0}
      \frac{\mathbb{P}\!\left(\text{down-crossing in }
            [t,\,t+\Delta t] \;\mid\; \GPField[t] > 0\right)}{\Delta t} \notag \\
    &= \frac{\displaystyle\lim_{\Delta t \to 0}
        \frac{\mathbb{P}\left(\text{down-crossing in }
               [t,\,t+\Delta t] \,\cap\, \GPField[t] > 0\right)}{\Delta t}}
      {\mathbb{P}(\GPField[t] > 0)} \notag \\
    \label{eqn:local_conditioning}
    &= \frac{\DownCross[\mathbf{r}(t),\RayDir]}
            {\normCDF[\MeanF[t]/\StdF[t]]}.
\end{align}
Here $\DownCross[\mathbf{r}(t),\RayDir]$ is the
\emph{down-crossing density}: the expected number of down-crossings
per unit ray length. Dividing by
$\mathbb{P}(\GPField[t]>0) = \normCDF[\MeanF[t]/\StdF[t]]$
converts this ensemble rate into one conditioned only on the current
point being in empty space. Throughout the remainder of the paper, we use
$\ExtCoeff[\mathbf{r}(t),\RayDir]$ for the local extinction coefficient
defined in \cref{eqn:local_conditioning}, while $\ExtCoeffExact[t]$
denotes the exact, history-dependent extinction rate.

The same local approximation has been used for
Gaussian height fields and level sets of 3D Gaussian fields with
spatially uniform roughness statistics
\citep{Lumme:1990:Diffuse,Peltoniemi:1993:Radiative}.

\subsection{Down-Crossing Density and One-Sided Projected Area}
\label{sec:kacrice}

The Kac--Rice formula
\citep{Kac:1943:Average,rice1945mathematical}
expresses the down-crossing density introduced in
\cref{eqn:local_conditioning} as an integral over directional derivative
values:
\begin{align}
  \label{eqn:downcrossing}
  \DownCross[\mathbf{r}(t),\RayDir] = \int_{-\infty}^{0} |v|\,
    p_{\GPField[t],\DirDeriv[t]}(0,\, v)\, \dif v.
\end{align}
Here, $v$ denotes a possible value of the directional derivative at a zero crossing. Restricting the integral to $v<0$ selects down-crossings, while the factor $|v|$ accounts for steeper crossings being encountered more frequently.

Factoring the joint PDF into a marginal and conditional density,
$p_{\GPField[t],\DirDeriv[t]}(0,v)
= p_{\GPField[t]}(0)\,p_{\CondDeriv[t]}(v)$, gives:
\begin{align}
  \label{eqn:downcrossing_factored}
  \DownCross[\mathbf{r}(t),\RayDir]
    &= p_{\GPField[t]}(0)
      \int_{-\infty}^{0} |v|\, p_{\CondDeriv[t]}(v)\, \dif v,\quad \text{where}
\end{align}
\begin{align}
    p_{\GPField[t]}(0) &= \frac{\normPDF[\MeanF[t]/\StdF[t]]}{\StdF[t]}, \;\;
    p_{\CondDeriv[t]}(v) = \mathcal{N}(v;\, \CondDerivMean[t],\,\CondDerivVar[t])
\end{align}
with the conditional derivative statistics given by \cref{eqn:ray_cond_stats}.

The two factors in \cref{eqn:downcrossing_factored} play distinct roles. The marginal 
density $p_{\GPField[t]}(0)$ measures how much field probability is concentrated near 
zero at the current point. The remaining integral is the expected magnitude of the 
descending directional derivatives at the zero set, measuring how strongly the 
surface faces the ray. We therefore call 
this quantity the \emph{one-sided projected area} $\ProjAreaMinus[\mathbf{r}(t),
\RayDir]$. Its Gaussian integral evaluates in closed form:
\begin{keyboxtitled}{blue!60!black}{One-Sided Projected Area}
\begin{subequations}
\label{eqn:one_sided_area}
\begin{align}
\ProjAreaMinus[\mathbf{r}(t),\RayDir]
    \coloneq&{}
    \int_{-\infty}^{0} |v|\, p_{\CondDeriv[t]}(v)\, \dif v
    \label{eqn:one_sided_area_def}\\
    =&{\;}
    \CondDerivStd[t]\,\normPDF[\frac{\CondDerivMean[t]}{\CondDerivStd[t]}]
    - \CondDerivMean[t]\,\normCDF[-\frac{\CondDerivMean[t]}{\CondDerivStd[t]}].
    \label{eqn:one_sided_area_closed}
\end{align}
\end{subequations}
\end{keyboxtitled}
\noindent Substituting this into
\cref{eqn:downcrossing_factored} yields:
\begin{align}
  \label{eqn:kacrice_density}
  \DownCross[\mathbf{r}(t),\RayDir]
    = p_{\GPField[t]}(0)\,
      \ProjAreaMinus[\mathbf{r}(t),\RayDir].
\end{align}
In \cref{sec:ndfderivation}, we show that the
same $\ProjAreaMinus$ is also the projected area of the NDF
\eqref{eqn:projarea_def} and the normalization of the vNDF
\eqref{eqn:vndf}. Thus, $\ProjAreaMinus$ connects the local crossing
rate derived here to the distribution of normals encountered at
scattering events.

\Cref{fig:kac_rice_bunny} visualizes the factorization in
\cref{eqn:kacrice_density} for a 2D GPIS. Realizations of the zero set
wiggle in a band around the mean surface, and the marginal density
$p_{\GPField}(0)$ identifies this band. Weighting it by the one-sided
projected area from \cref{eqn:one_sided_area_def} suppresses portions
of the surface that do not face the ray, producing the directional
down-crossing density $\DownCross$. In more uncertain regions, 
both the band and the crossing density broaden.

\subsection{Extinction, Transmittance, and Free-Flight PDF}
\label{sec:transmittance}

Substituting \cref{eqn:kacrice_density} into \cref{eqn:local_conditioning} gives
the closed-form extinction coefficient under the local-conditioning approximation:
\begin{keyboxtitled}{blue!60!black}{Extinction Coefficient}
\begin{align}
  \label{eqn:extinction_coeff}
  \ExtCoeff[\mathbf{r}(t),\RayDir]
    =
    \mymathbox{orange!70!black}{occupancy}{%
      \dfrac{\normPDF[\MeanF[t]/\StdF[t]]}{\StdF[t]\;\normCDF[\MeanF[t]/\StdF[t]]}%
      }%
    \;
    \mymathbox{blue!60!black}{directional proj.\ area}{%
      \ProjAreaMinus[\mathbf{r}(t),\RayDir]
      }%
      \!\!\!.
\end{align}
\end{keyboxtitled}

\noindent The occupancy factor depends only on the field's marginal
statistics at $\mathbf{r}(t)$: it is large near probable surfaces
($\MeanF/\StdF$ near zero), small deep in empty space, and continues
to grow inside the solid, as illustrated in the last two columns of
\cref{fig:kac_rice_bunny}. The projected-area factor
$\ProjAreaMinus[\mathbf{r}(t),\RayDir]$ contains all dependence on
$\RayDir$ through the conditional derivative statistics. It becomes direction-dependent whenever
$\CondGrad[\mathbf{r}(t)]$ has a non-zero mean or anisotropic covariance,
yielding the extinction of an anisotropic RTE
\citep{Jakob:2010:Radiative}. When $\CondGradMean=\mathbf{0}$ and
$\CondGradCov\propto\mathbf{I}$, the extinction is isotropic.

Replacing the exact rate $\ExtCoeffExact$ in
\cref{eqn:transmittance_ode} with the local coefficient from
\cref{eqn:extinction_coeff} and integrating gives
\begin{align}
  \label{eqn:transmittance}
  \Trans[t] = \exp\!\left( -\int_0^t \ExtCoeff[s]\, \dif s \right)
\end{align}
as the local approximation to $\TransExact[t]$, the probability that the ray reaches depth $t$ without a down-crossing.

Together, \cref{eqn:extinction_coeff,eqn:transmittance} complete the
collision-distance law of \cref{eq:re}: the free-flight density
$\FreeFlightPDF[t] = \ExtCoeff[t]\,\Trans[t]$ carries total mass
$1 - \Trans[\infty]$, and the remainder $\Trans[\infty]$ is the
probability that the ray escapes to the background.

\subsection{The Normal Distribution Function}
\label{sec:ndfderivation}

The collision-distance law in
\cref{eqn:extinction_coeff,eqn:transmittance} determines \emph{where} a ray
scatters. To determine \emph{how} it scatters, we need the distribution of
surface normals encountered at those down-crossings. For the remainder
of this section, we fix $\mathbf{x}$ and omit it
from the notation.

From \cref{eqn:grad_cond_stats}, the gradient conditioned on a zero
field value has density
$p_{\CondGrad}(\GradField)
\coloneq
\mathcal{N}(\GradField;\,\CondGradMean,\,\CondGradCov)$,
where $\GradField\in\mathbb{R}^3$ is a possible gradient value. But this
pointwise gradient distribution is not the distribution observed by a
ray. Two effects reweight it:
\begin{itemize}
  \item \emph{Crossing-rate weighting:}
  By \cref{eqn:downcrossing}, a gradient $\GradField$ contributes
  crossings at a rate proportional to
  $|\GradField\cdot\RayDir|$, so steeper crossings are encountered
  more frequently per unit length.

  \item \emph{Down-crossing selection:}
  A ray traveling through empty space encounters the zero set only
  when $\GradField\cdot\RayDir<0$.
\end{itemize}
Together, these effects give the following unnormalized density for
gradients encountered at down-crossings:
\begin{align}
  \label{eqn:crossing_gradient_law}
  p\!\left(
    \GradField
    \mid \text{down-crossing along }\RayDir
  \right)
  \propto
  \langle-\RayDir,\GradField\rangle\,
  p_{\CondGrad}(\GradField).
\end{align}
\Cref{eqn:crossing_gradient_law} describes a typical down-crossing.
The gradient distribution at a \emph{first} intersection would
additionally condition on the absence of earlier crossings,
information discarded by the local-conditioning approximation.

To obtain the distribution of normals, we write
$\GradField=r\SurfNorm$, where $r=\|\GradField\|>0$.
Under this change of variables, the volume element becomes
$\dif\GradField=r^2\,\dif r\,\dif\SurfNorm$, while the crossing-rate
weight becomes
$\langle-\RayDir,\GradField\rangle
=r\,\langle-\RayDir,\SurfNorm\rangle$.
The spherical Jacobian and crossing-rate weight therefore contribute
the combined factor $r^3\,\langle-\RayDir,\SurfNorm\rangle$.\footnote{The projected normal (angular Gaussian) distribution has
radial weight $r^2$; the extra $r$ here comes from Kac--Rice
crossing-rate weighting.}
Marginalizing over the gradient magnitude $r$ then gives the
normal density up to normalization:
\begin{align}
  \label{eqn:crossing_normal_density}
  p\!\left(
    \SurfNorm
    \mid \text{down-crossing along }\RayDir
  \right)
  \propto
  \langle-\RayDir,\SurfNorm\rangle
  \int_0^\infty p_{\CondGrad}(r\SurfNorm)\,r^3\,\dif r.
\end{align}

The radial integral in \cref{eqn:crossing_normal_density} collects
contributions from gradients of all magnitudes aligned with
$\SurfNorm$. It characterizes the normal orientations independently
of the ray direction $\RayDir$, so we define it as the
\emph{normal distribution function} (NDF):\footnote{We use NDF in the microfacet and microflake sense of an
oriented surface-area density, rather than a probability density.}
\begin{align}
  \label{eqn:ndf_def}
  \NDF[\SurfNorm]
    \coloneq
    \int_0^\infty p_{\CondGrad}(r\SurfNorm)\,r^3\,\dif r.
\end{align}
Although $\NDF$ need not integrate to one, it has a direct geometric
meaning:
$p_{\GPField}(0)\,\NDF[\SurfNorm]\,\dif\SurfNorm$
is the expected zero-set area per unit volume whose oriented normal
lies in $\dif\SurfNorm$.\footnote{Integrating over all directions gives
the total expected zero-set area per unit volume.}

Integrating the unnormalized density in
\cref{eqn:crossing_normal_density} over all normal directions gives
\begin{align}
  \label{eqn:projarea_def}
  \int_{S^2}
    \langle-\RayDir,\SurfNorm\rangle\,
    \NDF[\SurfNorm]\,\dif\SurfNorm
  &=
  \int_{\mathbb{R}^3}
    \langle-\RayDir,\GradField\rangle\,
    p_{\CondGrad}(\GradField)\,\dif\GradField
    \notag\\
  &=
  \int_{-\infty}^{0}
    |v|\,p_{\CondDeriv}(v)\,\dif v
  =
  \ProjAreaMinus[\RayDir].
\end{align}
The final expression is precisely the one-sided projected area from
\cref{eqn:one_sided_area_def}, with the fixed position omitted.
Thus, the same projected area that determines the local down-crossing
rate in \cref{eqn:kacrice_density} also normalizes the distribution
of normals encountered by the ray.

Dividing the unnormalized density in
\cref{eqn:crossing_normal_density} by
$\ProjAreaMinus[\RayDir]$ thus gives the visible normal distribution
function (vNDF):
\begin{keyboxtitled}{blue!60!black}{Visible Normal Distribution Function (vNDF)}
\begin{align}
  \label{eqn:vndf}
  \VNDF[\RayDir][\SurfNorm]
    =
    \frac{
      \langle-\RayDir,\SurfNorm\rangle\,
      \NDF[\SurfNorm]
    }{
      \ProjAreaMinus[\RayDir]
    }.
\end{align}
\end{keyboxtitled}

\Cref{eqn:vndf} has the standard visible-normal form from microfacet
and microflake theory: the NDF is weighted by a clamped cosine and
normalized by its projected area
\citep{Jakob:2010:Radiative,Heitz:2015:SGGX}. In classical models, the
NDF is a modeling input; here, both the NDF and its projected area are
induced by the GPIS statistics.

\paragraph{Evaluating the NDF}
\label{sec:ndfevaluation}

To evaluate $\NDF[\SurfNorm]$ in closed form, we substitute
$p_{\CondGrad} = \mathcal{N}(\CondGradMean,\,\CondGradCov)$
into \cref{eqn:ndf_def} and expand the Gaussian exponent
at $\GradField = r\SurfNorm$:
\begin{align}
  -\tfrac{1}{2}(r\SurfNorm - \CondGradMean)^\mathsf{T}
    \CondGradCov^{-1}
    (r\SurfNorm - \CondGradMean)
  = -\tfrac{1}{2}\!\left(\NDFA[\SurfNorm]\,r^2 - 2\NDFB[\SurfNorm]\,r + \NDFC\right),
\end{align}
where we define three scalar quantities:
\begin{align}
  \label{eqn:ndf_abc}
  \NDFA[\SurfNorm] \coloneq \SurfNorm^\mathsf{T}\CondGradCov^{-1}\SurfNorm, \quad
  \NDFB[\SurfNorm] \coloneq \SurfNorm^\mathsf{T}\CondGradCov^{-1}\CondGradMean, \quad
  \NDFC \coloneq \CondGradMean[][\mathsf{T}]\CondGradCov^{-1}\CondGradMean.
\end{align}
$\NDFA[\SurfNorm] > 0$ for all $\SurfNorm$ since $\CondGradCov^{-1}$ is positive
definite, and $\NDFC$ is a constant independent of $\SurfNorm$.
With the shorthand $\NDFb[] \coloneq \NDFB[\SurfNorm]/\sqrt{\NDFA[\SurfNorm]}$
(dependence on $\SurfNorm$ suppressed for brevity), the radial integral of
\cref{eqn:ndf_def} evaluates analytically, as shown in \cref{app:ndfintegral}:
\begin{keyboxtitled}{blue!60!black}{Normal Distribution Function (NDF)}%
\begin{align}%
  \label{eq:ndf}
  \NDF[\SurfNorm] =
    \frac{\euler^{(\NDFb[][2] - \NDFC)/2}}
         {2\pi\,|\CondGradCov|^{1/2}\,\NDFA[]^2}
    \left[
      \left(\NDFb[][2]+2\right)
        \normPDF[\NDFb[]] + \NDFb[]\left(\NDFb[][2]+3\right)\normCDF[\NDFb[]]
    \right].
\end{align}%
\end{keyboxtitled}%

\begin{figure}[t]
  \centering
  \setlength{\tabcolsep}{1pt}
  \renewcommand{\arraystretch}{0.5}
  \begin{tabular}{@{}ccccc@{}}
    \makebox[0.2\linewidth][c]{\footnotesize SGGX} &
    \makebox[0.2\linewidth][c]{} &
    \makebox[0.2\linewidth][c]{} &
    \makebox[0.2\linewidth][c]{} &
    \makebox[0.2\linewidth][c]{\footnotesize Beckmann} \\[2pt]
    \includegraphics[width=0.2\linewidth]{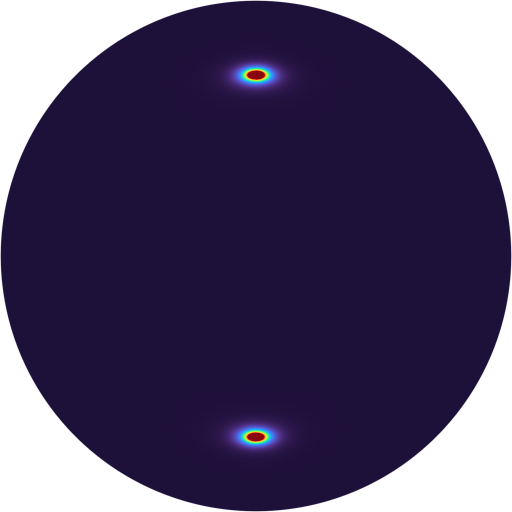} &
    \includegraphics[width=0.2\linewidth]{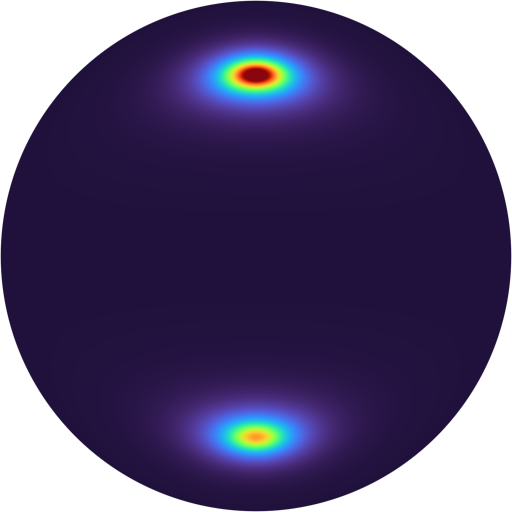} &
    \includegraphics[width=0.2\linewidth]{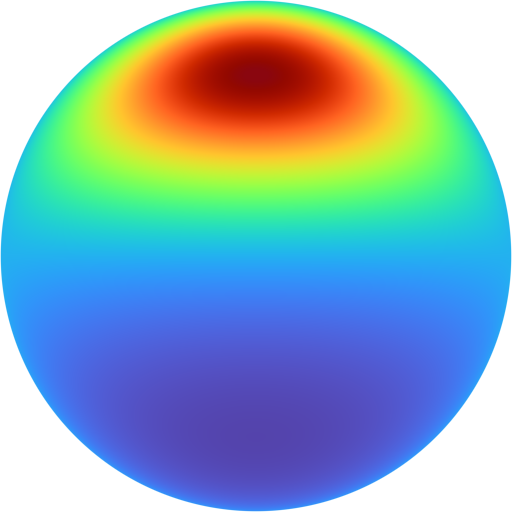} &
    \includegraphics[width=0.2\linewidth]{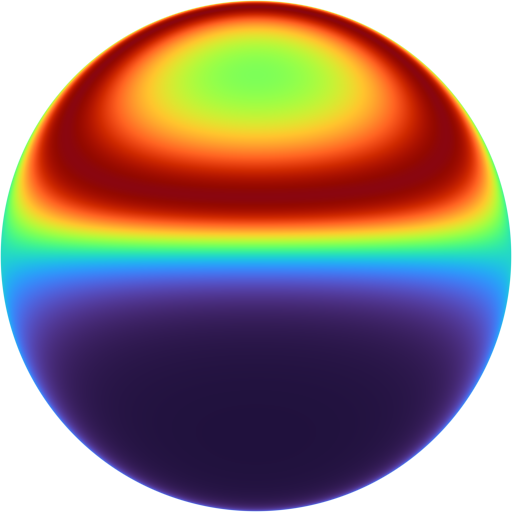} &
    \includegraphics[width=0.2\linewidth]{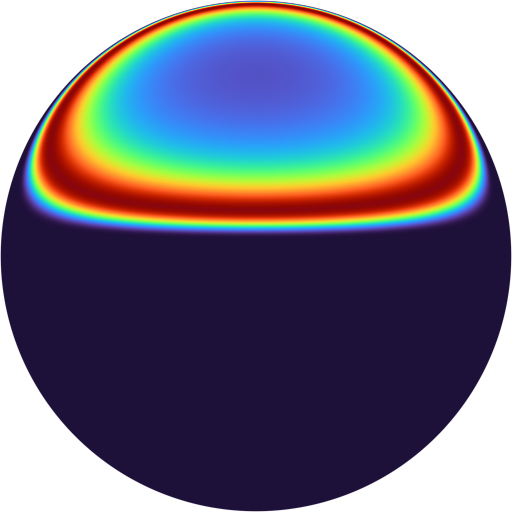} \\[2pt]
    \makebox[0.2\linewidth][c]{\footnotesize $\ell_z = 0.05$} &
    \makebox[0.2\linewidth][c]{\footnotesize $\ell_z = 0.2$} &
    \makebox[0.2\linewidth][c]{\footnotesize $\ell_z = 1$} &
    \makebox[0.2\linewidth][c]{\footnotesize $\ell_z = 2$} &
    \makebox[0.2\linewidth][c]{\footnotesize $\ell_z = 100$} \\
  \end{tabular}
  \caption{NDF as a function of the GP correlation length scale $\ell_z$, with fixed roughness $\alpha = 2.0$ and mean gradient $\partial\mu/\partial z = 1.0$. At small $\ell_z$ the distribution closely matches SGGX (far left); as $\ell_z$ grows it converges to the Beckmann NDF (far right).}
  \label{fig:lz-sweep}
\end{figure}

\begin{figure}[t]
  \centering
  \setlength{\tabcolsep}{1pt}
  \renewcommand{\arraystretch}{0.5}
  \begin{tabular}{@{}r@{\hspace{1pt}}cccc@{}}
    \rotatebox{90}{\makebox[0.23\linewidth][c]{\footnotesize $\ell_z = 0.5$}} &
    \includegraphics[width=0.23\linewidth]{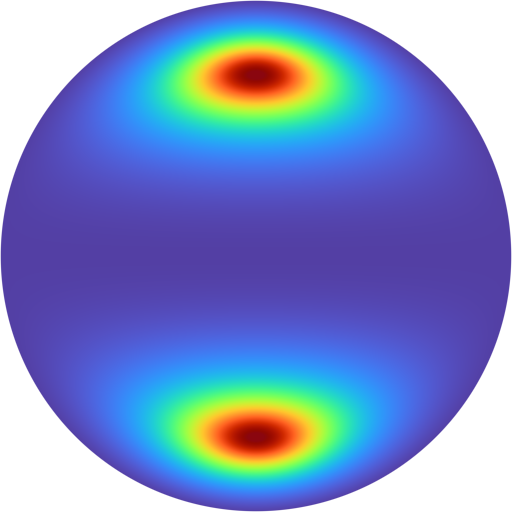} &
    \includegraphics[width=0.23\linewidth]{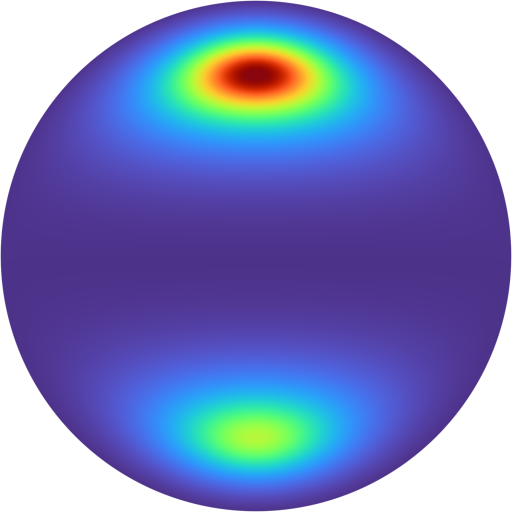} &
    \includegraphics[width=0.23\linewidth]{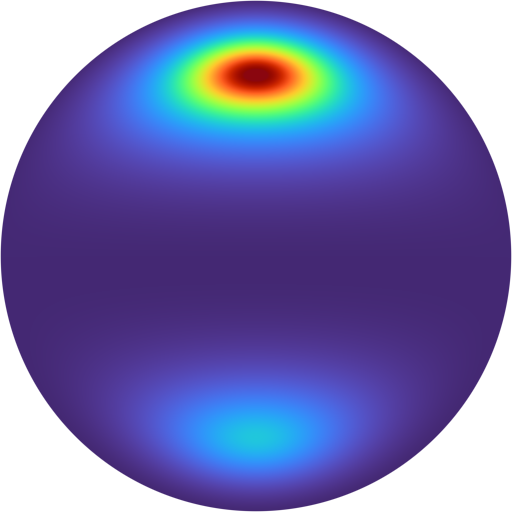} &
    \includegraphics[width=0.23\linewidth]{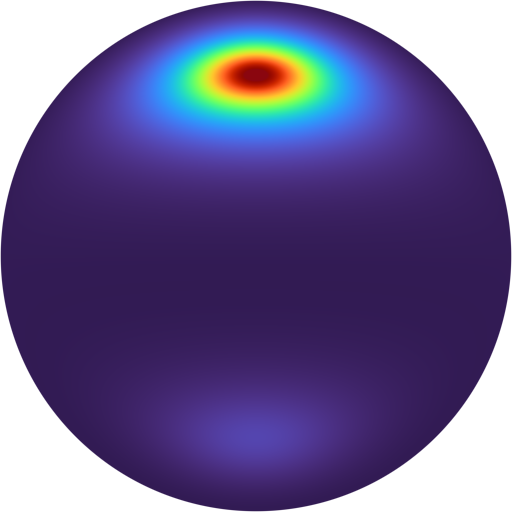} \\
    \rotatebox{90}{\makebox[0.23\linewidth][c]{\footnotesize $\ell_z = 2.0$}} &
    \includegraphics[width=0.23\linewidth]{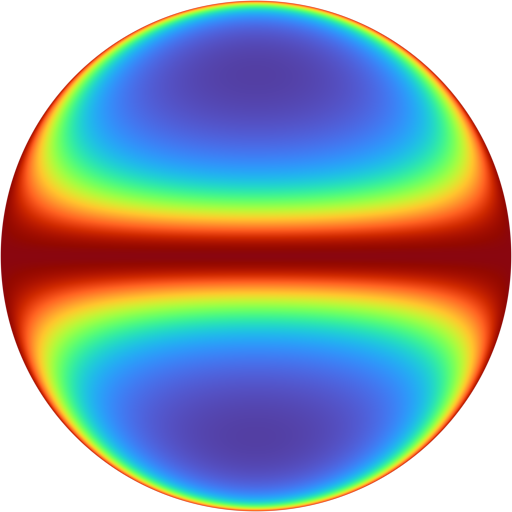} &
    \includegraphics[width=0.23\linewidth]{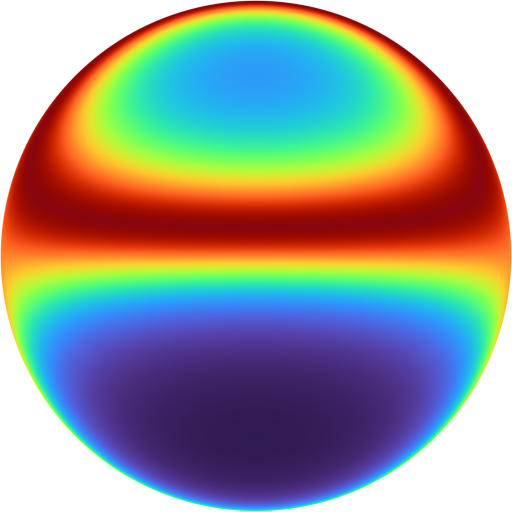} &
    \includegraphics[width=0.23\linewidth]{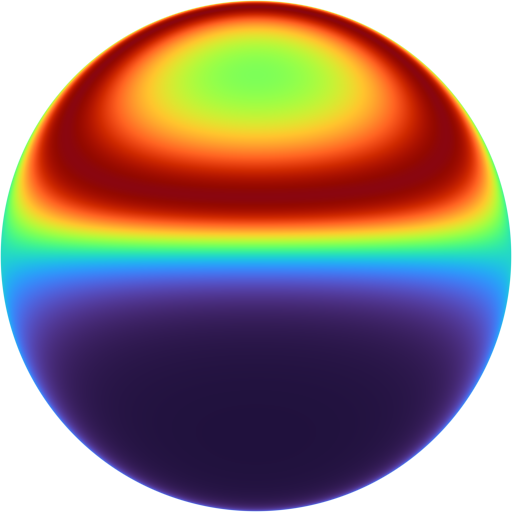} &
    \includegraphics[width=0.23\linewidth]{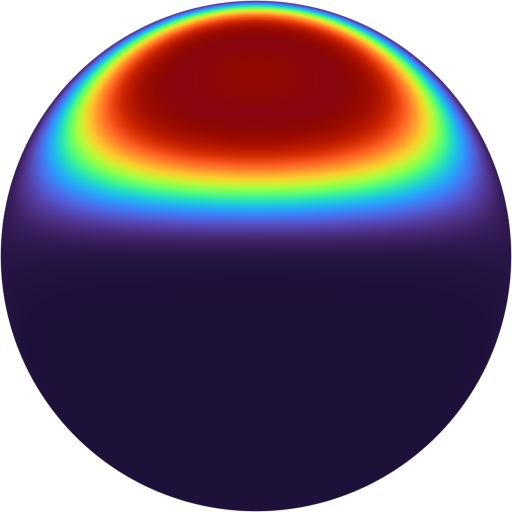} \\[2pt]
     &
    \makebox[0.23\linewidth][c]{\footnotesize $\partial\mu/\partial z = 0$} &
    \makebox[0.23\linewidth][c]{\footnotesize $\partial\mu/\partial z = 0.5$} &
    \makebox[0.23\linewidth][c]{\footnotesize $\partial\mu/\partial z = 1$} &
    \makebox[0.23\linewidth][c]{\footnotesize $\partial\mu/\partial z = 2$} \\
  \end{tabular}
  \caption{NDF as a function of the mean gradient magnitude $\partial\mu/\partial z$ (columns), for two values of the GP correlation length scale $\ell_z$ (rows), with fixed roughness $\alpha = 2.0$. At $\partial\mu/\partial z = 0$ the distribution is symmetric (equivalent to SGGX); increasing $\partial\mu/\partial z$ progressively breaks symmetry.}
  \label{fig:mgz-sweep}
\end{figure}

\Cref{fig:lz-sweep,fig:mgz-sweep} illustrate how the correlation
length and mean gradient shape the NDF, using mirrored-sphere
renderings to show the full spherical distribution.
We set $\CondGradMean=(0,0,\CondGradMag)^\mathsf{T}$ and
$\CondGradCov=(\alpha^2/2)\operatorname{diag}(1,1,\CorLen_z^{-2})$.
Here, $\CorLen_z$ is the GP correlation length along $z$, with
unit correlation lengths along $x$ and $y$.

At small $\CorLen_z$, the normals span the full sphere and the NDF
approaches SGGX; as $\CorLen_z$ increases, they concentrate around
the mean-gradient direction and the NDF approaches Beckmann
(\cref{fig:lz-sweep}). The mean gradient controls its asymmetry
(\cref{fig:mgz-sweep}): when $\CondGradMag=0$, the NDF is antipodally
symmetric, while increasing $\CondGradMag$ biases the normals
toward $+z$. \Cref{sec:ndfproperties} analyzes these limiting cases.

\begin{figure}[t]
  \centering
  \setlength{\tabcolsep}{1pt}
  \renewcommand{\arraystretch}{0.5}
  \begin{tabular}{@{}r@{\hspace{1pt}}cccc@{}}
    \rotatebox{90}{\makebox[0.23\linewidth][c]{\footnotesize $(2, 0.5, 0)$}} &
    \includegraphics[width=0.23\linewidth]{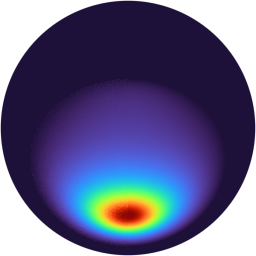} &
    \includegraphics[width=0.23\linewidth]{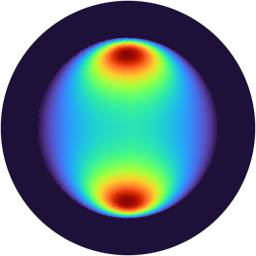} &
    \includegraphics[width=0.23\linewidth]{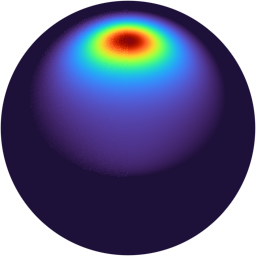} &
    \includegraphics[width=0.23\linewidth]{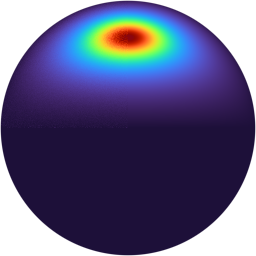} \\
    \rotatebox{90}{\makebox[0.23\linewidth][c]{\footnotesize $(1, 2, 1)$}} &
    \includegraphics[width=0.23\linewidth]{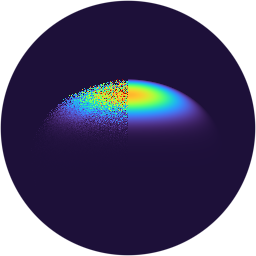} &
    \includegraphics[width=0.23\linewidth]{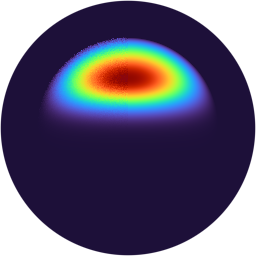} &
    \includegraphics[width=0.23\linewidth]{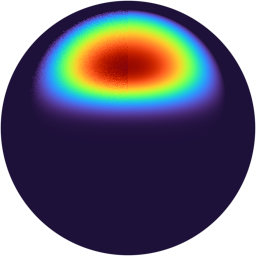} &
    \includegraphics[width=0.23\linewidth]{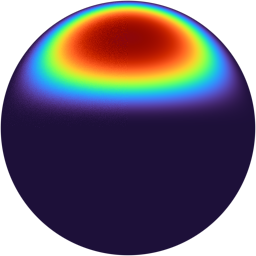} \\[2pt]
     &
    \makebox[0.23\linewidth][c]{\footnotesize $\theta = -45^\circ$} &
    \makebox[0.23\linewidth][c]{\footnotesize $\theta = 0^\circ$} &
    \makebox[0.23\linewidth][c]{\footnotesize $\theta = 45^\circ$} &
    \makebox[0.23\linewidth][c]{\footnotesize $\theta = 90^\circ$} \\
  \end{tabular}
  \caption{Analytic GPIS vNDF (right half of each panel) versus ray-marched reference
(left half, 4096~spp) across four incident angles (columns;
$\theta=90^\circ$ normal, $\theta=0^\circ$ grazing) and
physical parameter settings (rows).
Panels are normalised independently.}
  \label{fig:vndf-rm-vs-analytic}
\end{figure}

\Cref{fig:vndf-rm-vs-analytic} validates the analytic vNDF obtained
from \cref{eq:ndf,eqn:vndf} against Monte Carlo estimates from
ray-marched GPIS realizations. Across both parameter configurations
and all four incidence angles, the analytic result on the right half
of each panel matches the ray-marched estimate on the left, including
at grazing incidence.

\subsection{The Phase Function}
\label{sec:phasefunction}

At a down-crossing, the surface normal follows the vNDF in
\cref{eqn:vndf}. For a given normal $\SurfNorm$, the micro-BRDF
$f_r(-\RayDir,\ScatDir;\SurfNorm)$ determines scattering into
direction $\ScatDir$. Averaging over the vNDF then gives the standard microflake phase function
\citep{Jakob:2010:Radiative}

\begin{align}
  \label{eqn:phasefn}
  \PhaseFunc[\RayDir][\ScatDir]
    = \int_{S^2} f_r(-\RayDir,\, \ScatDir;\, \SurfNorm)\,
      \langle\ScatDir,\SurfNorm\rangle\, \VNDF[\RayDir][\SurfNorm]\, \dif\SurfNorm.
\end{align}
For a perfect mirror micro-BRDF, the delta function selects the
half-vector $\HalfVec=(\ScatDir-\RayDir)/\|\ScatDir-\RayDir\|$.
Applying the reflection Jacobian gives~\citep{Jakob:2010:Radiative,
Heitz:2015:SGGX}:
\begin{align}
  \label{eq:mirrorpf}
  \PhaseFunc[\RayDir][\ScatDir]
    = \frac{\VNDF[\RayDir][\HalfVec]}{4|\RayDir\cdot\HalfVec|}
    = \frac{\NDF[\HalfVec]}{4\,\ProjAreaMinus[\RayDir]}.
\end{align}
The phase function integrates to one over $\ScatDir$ for a
nonabsorbing micro-BRDF and to less than one with absorption.

\subsection{Properties and Limits}
\label{sec:ndfproperties}

We examine the full-sphere support and asymmetry of the GPIS NDF,
together with the reciprocity of the induced transport. We show
that the centered case recovers SGGX exactly and how the one-sided
construction encompasses both asymmetric and symmetric microflake
transport. The next section (\cref{sec:surfaceconnections}) develops surface
models with full-sphere NDFs and recovers Beckmann--Smith scattering
in the height-field limit.

\paragraph{Full-sphere support}
Unlike hemispherical NDFs in classical microfacet theory,
$\NDF[\SurfNorm]$ has support on the entire sphere, accommodating
downward-pointing normals from overhangs and re-entrant geometry
that height fields cannot represent.

\paragraph{Asymmetry and one-sided projected areas}
Unlike symmetric microflake NDFs such as SGGX, our NDF is generally
asymmetric:
$\NDF[\SurfNorm]\neq\NDF[-\SurfNorm]$.
We therefore define the complementary one-sided projected area
$\ProjAreaPlus[\RayDir]$ over normals facing along $\RayDir$.
Equivalently,
$\ProjAreaPlus[\RayDir]=\ProjAreaMinus[-\RayDir]$ is the down-crossing
projected area for the reversed ray direction:
\begin{subequations}
\label{eqn:projarea_plus}
\begin{align}
  \ProjAreaPlus[\RayDir]
    &\coloneq
    \int_{S^2}\!
      \langle\RayDir,\SurfNorm\rangle\,\NDF[\SurfNorm]\,\dif\SurfNorm
    \label{eqn:projarea_plus_def}\\
    &= \CondDerivStd\,\normPDF[\frac{\CondDerivMean}{\CondDerivStd}]
      + \CondDerivMean\,\normCDF[\frac{\CondDerivMean}{\CondDerivStd}].
    \label{eqn:projarea_plus_closed}
\end{align}
\end{subequations}
The vector first moment of the NDF is the conditional mean gradient:
substituting $\GradField = r\SurfNorm$ in \cref{eqn:ndf_def} gives
\begin{align}
  \label{eqn:ndf_first_moment}
  \int_{S^2} \SurfNorm\,\NDF[\SurfNorm]\,\dif\SurfNorm
    = \CondGradMean.
\end{align}
Taking the dot product with $\RayDir$ gives:
\begin{align}
  \label{eqn:asymmetry}
  \ProjAreaPlus[\RayDir] - \ProjAreaMinus[\RayDir]
    = \CondDerivMean
    = \CondGradMean\cdot\RayDir.
\end{align}
The difference between the two areas is exactly the projection of
the conditional mean gradient onto the ray; it vanishes precisely
when $\CondDerivMean=0$.

\paragraph{Reciprocity}
Because $\ProjAreaMinus[\RayDir]$ depends on direction, simple phase
function symmetry $\PhaseFunc[\RayDir][\ScatDir] =
\PhaseFunc[\ScatDir][\RayDir]$ does not hold in general, as discussed 
for one-sided height-field
media by \citet{Dupuy:2016:Additional}.
For a reciprocal micro-BRDF, reciprocity
instead relates each scattering event to its reverse, traversed in the opposite
direction
\begin{align}
  \label{eqn:weighted_reciprocity}
  \ProjAreaMinus[\RayDir]\, \PhaseFunc[\RayDir][\ScatDir]
    = \ProjAreaMinus[-\ScatDir]\, \PhaseFunc[-\ScatDir][-\RayDir],
\end{align}
which our mirror phase function \eqref{eq:mirrorpf} satisfies exactly:
the reverse path shares the half-vector $\HalfVec$, so both sides
equal $\NDF[\HalfVec]/4$. 

\paragraph{Centered case and high roughness limit: SGGX}
\label{app:sggxlimit}
When $\CondGradMean = \mathbf{0}$, the scalars $\NDFB[\SurfNorm]$ and
$\NDFC$ from \cref{eqn:ndf_abc} and the ratio $\NDFb[]$ vanish.
Substituting into \cref{eq:ndf} with $\normPDF[0] = 1/\sqrt{2\pi}$
gives:
\begin{align}
  \label{eqn:ndf_zeromean}
  \NDF[\SurfNorm]\big|_{\CondGradMean=\mathbf{0}}
    = \frac{1}{\pi\sqrt{2\pi}\,|\CondGradCov|^{1/2}
      \big(\SurfNorm^\mathsf{T}\CondGradCov^{-1}\SurfNorm\big)^2}.
\end{align}
This is exactly the SGGX NDF~\citep{Heitz:2015:SGGX}
$D_\mathrm{SGGX}(\SurfNorm) =
1/(\pi|\SGGXMat|^{1/2}(\SurfNorm^\mathsf{T}\SGGXMat^{-1}\SurfNorm)^2)$
with $\SGGXMat = \CondGradCov/(2\pi)$.
The centered case is equally the high-roughness limit: holding
$\CondGradMean$ fixed while the conditional gradient spread grows in
all directions, $\NDFC \to 0$ and the NDF converges to the SGGX form
above.
Furthermore, $\CondDerivMean = 0$ and $\CondDerivStd^2 =
\RayDir^\mathsf{T}\CondGradCov\,\RayDir = 2\pi\,\RayDir^\mathsf{T}\SGGXMat\,\RayDir$, so
\cref{eqn:one_sided_area,eqn:projarea_plus} both recover the SGGX
projected area exactly:
$\ProjAreaMinus[\RayDir] = \ProjAreaPlus[\RayDir] =
\CondDerivStd\,\normPDF[0] = \sqrt{\RayDir^\mathsf{T}\SGGXMat\,\RayDir}
\eqcolon a_\mathrm{SGGX}(\RayDir)$.

\paragraph{Unifying symmetric and asymmetric microflake transport}
One-sidedness follows from the down-crossing condition for rays
in empty space, recovering the microflake interpretation of
height-field models
\citep{Heitz:2016:Multiplescattering,Dupuy:2016:Additional}.
This encounter rule applies to both asymmetric and symmetric NDFs.
In the centered case $\CondGradMean=\mathbf{0}$, opposite normal
orientations have equal density and the NDF reduces to SGGX.
With the same material law for opposite orientations, this recovers
conventional two-sided microflake transport.
Both regimes therefore follow from the same GPIS crossing
statistics, without introducing separate one-sided and two-sided flake models.

\subsection{Heavy-Tailed Extension via Scale Mixtures}
\label{sec:mixture}

The distribution of gradients of any GPIS will be Gaussian by definition.
However, we can still obtain heavy-tailed NDF variants analogous to GGX
by applying a scale mixture to the zero-conditioned gradient
law~\citep{dEon:2023:StudentT}: draw
$\tau \sim \mathrm{Exp}(1)$ and replace
$\CondGradCov \to \CondGradCov/\tau$ in the conditional gradient density
of \cref{sec:ndfderivation}, then marginalize over $\tau$.
Substituting the resulting heavy-tailed gradient density into the
Kac--Rice radial integral \eqref{eqn:ndf_def} and evaluating in closed
form (\cref{app:mixture}) gives the heavy-tailed NDF
\begin{keyboxtitled}{blue!60!black}{Heavy-Tailed NDF}
\begin{align}
  \label{eqn:mixture_ndf}
  \!\!\!\NDF[\SurfNorm]^{\mathrm{GGX}}
    &= \frac{2S-\mu}{2\pi\,|\CondGradCov|^{1/2}\,\NDFA[]^{5/2}\,(S-\mu)^2},\;
  \mu \coloneq \tfrac{\NDFB[]}{\NDFA[]}, \;
  S \coloneq \sqrt{\tfrac{\NDFC+2}{\NDFA[]}},\!\!
\end{align}
\end{keyboxtitled}
\noindent with $\NDFA,\NDFB, \NDFC$ from \cref{eqn:ndf_abc}.
The same mixture applied one dimension down to the ray-projected marginal
(\cref{app:mixture}) gives the one-sided projected area, with
$\alpha_{\mathrm{eff}}^2 \coloneq 2\,\CondDerivVar$:
\begin{keyboxtitled}{blue!60!black}{Heavy-Tailed One-Sided Projected Area}
\begin{align}
  \label{eqn:mixture_projarea}
  \ProjAreaMinus[\RayDir]^{\mathrm{GGX}}
    &= \frac{\sqrt{\CondDerivMean^2+\alpha_{\mathrm{eff}}^2}-\CondDerivMean}{2},
\end{align}
\end{keyboxtitled}
\noindent and $\ProjAreaPlus[\RayDir]^{\mathrm{GGX}} = \ProjAreaMinus[\RayDir]^{\mathrm{GGX}} + \CondDerivMean$. In the height-field limit ($\CondGradCov=\mathrm{diag}(\epsilon_h,\epsilon_h,\epsilon_z)$,
$\epsilon_z\to0$), the mixture commutes with the limit and exactly recovers the classical Trowbridge–Reitz/GGX
NDF.

\section{Surface Scattering from GPISes}
\label{sec:surfaceconnections}

Beyond volume rendering, we can also leverage our transport framework to construct novel surface scattering models.

\subsection{BRDF-Compatible GPISes}
\label{sec:surfacelimit}

We say a GPIS is \emph{BRDF-compatible} if, within a local planar
patch, its mean field is a linear ramp $\GPMean[\mathbf{x}]=z$,
with $\StdF$ and $\CondGradCov$ constant. Here, $z$ measures height
above the mean surface. These conditions give
$\CondGradMean=\nabla\GPMean=\hat{\mathbf{z}}$, making the NDF
spatially constant while the occupancy factor depends only on
height. Integrating out lateral displacements yields a surface
BRDF depending only on the incident and outgoing directions.

This construction generalizes the classical height-field model.
The covariance $\CondGradCov$ may be full rank and contain
off-diagonal terms, allowing gradient fluctuations along depth
and correlations between depth and lateral components.
The zero set can therefore exhibit overhangs and porous or
re-entrant geometry.

For the heavy-tailed extension of \cref{sec:mixture}, we retain
the same occupancy factor and use the corresponding NDF and
projected area. Below, $\NDF$ and $\ProjAreaMinus$ refer to the
selected family.

\subsection{The Single-Scattering GPIS BRDF}
\label{sec:single-scattering-brdf}

Starting from the RTE \eqref{eq:re}, we derive the single-scattering BRDF of a
BRDF-compatible GPIS with a perfect mirror micro-BRDF.
The external directions $\OutDir[i],\OutDir[o]$ point
into the upper hemisphere $S_+^2$, with
$\cos\theta_\ast=\OutDir\cdot\hat{\mathbf{z}}>0$.
The NDF remains defined on the full sphere.

Single scattering uses unscattered background radiance
$\Radiance^{(0)}(z,\OutDir[i])
= \Trans[z\to\infty,\OutDir[i]]\,\RadianceBG[\OutDir[i]]$,
where the transmittance is the shadow-ray escape probability;
$\Trans[\infty\to z,-\OutDir[o]]$ denotes view-ray survival.

Substituting $\Radiance^{(0)}$ into \cref{eq:re} with
$\RayDir=-\OutDir[o]$ and $\dif z=-\cos\theta_o\,\dif t$,
then exchanging integrals, gives
\begin{align}
  \Radiance_o^{(1)}
  ={}& \frac{1}{\cos\theta_o}
  \int_{S_+^2}\int_{-\infty}^{\infty}
    \ExtCoeff[z,-\OutDir[o]]\,
    \Trans[\infty\to z,-\OutDir[o]]
    \notag\\
  &\quad{}\times
    \PhaseFunc[-\OutDir[o]][\OutDir[i]]\,
    \Trans[z\to\infty,\OutDir[i]]\,
    \dif z\,
    \RadianceBG[\OutDir[i]]\,
    \dif\OutDir[i].
  \label{eqn:brdf_single_radiance}
\end{align}
Compare this with the surface rendering equation:
\begin{equation}
  \Radiance_o^{(1)}
  = \int_{S_+^2}
    f_r^{(1)}(\OutDir[i],\OutDir[o])\,
    \RadianceBG[\OutDir[i]]\,
    \cos\theta_i\,\dif\OutDir[i].
  \label{eqn:brdf_single_surface_equation}
\end{equation}
We obtain the BRDF by matching the angular integrands
of the two expressions.

Writing the occupancy factor as
$\rho^{\mathrm{Smith}}(z)\coloneq
\normPDF[z/\StdF]/(\StdF\normCDF[z/\StdF])$ and
substituting this into \cref{eqn:extinction_coeff,eq:mirrorpf}
gives
\begin{align}
  \ExtCoeff[z,-\OutDir[o]]\,
  \PhaseFunc[-\OutDir[o]][\OutDir[i]]
  &=
  \rho^{\mathrm{Smith}}(z)\,
  \cancel{\ProjAreaMinus[-\OutDir[o]]}\,
  \frac{\NDF[\HalfVec]}
       {4\cancel{\ProjAreaMinus[-\OutDir[o]]}}
  \notag\\
  &= \rho^{\mathrm{Smith}}(z)\,
     \frac{\NDF[\HalfVec]}{4}.
  \label{eqn:brdf_extinction_phase_product}
\end{align}

Set $u=\normCDF[z/\StdF]$, the local probability of empty
space, so that
$\rho^{\mathrm{Smith}}(z)\,\dif z=\dif u/u$.
Define Smith's auxiliary function as
$\SmithLambda[\OutDir]\coloneq
\ProjAreaMinus[\OutDir]/\cos\theta_\ast$.
\Cref{app:heightfield-masking} gives its Gaussian and GGX
forms and derives the view and shadow transmittances:
\begin{equation}
  \Trans[\infty\to z,-\OutDir[o]]
  = u^{1+\SmithLambda[\OutDir[o]]},
  \qquad
  \Trans[z\to\infty,\OutDir[i]]
  = u^{\SmithLambda[\OutDir[i]]}.
  \label{eqn:brdf_height_transmittances}
\end{equation}

The $1/u$ from the occupancy factor cancels the extra power
of $u$ in the view transmittance.
Treating the two escape events as
conditionally independent at their shared crossing height,
the remaining height integral gives the joint
masking--shadowing factor:
\begin{equation}
  G_{12}(\OutDir[i],\OutDir[o])
  = \int_0^1
    u^{\SmithLambda[\OutDir[i]]+\SmithLambda[\OutDir[o]]}
    \,\dif u
  = \frac{1}
    {1+\SmithLambda[\OutDir[i]]+\SmithLambda[\OutDir[o]]}.
  \label{eqn:brdf_single_height_integral}
\end{equation}
Combining this with $\NDF[\HalfVec]/4$ and the two
cosine factors yields
\begin{keyboxtitled}{blue!60!black}{Single-Scattering GPIS BRDF}
\begin{equation}
  f_r^{(1)}(\OutDir[i],\OutDir[o])
  = \frac{\NDF[\HalfVec]\,
          G_{12}(\OutDir[i],\OutDir[o])}
         {4\cos\theta_i\cos\theta_o}.
  \label{eqn:brdf_single_full}
\end{equation}
\end{keyboxtitled}
For fixed conditional gradient statistics, $\StdF$ drops out
of the height integral, so the BRDF is independent of
transition thickness.

\paragraph{Height-field limit: Beckmann--Smith}
Taking
$\CondGradCov=\mathrm{diag}(\epsilon_h,\epsilon_h,\epsilon_z)$
and letting $\epsilon_z\to0$ with $\epsilon_h$ fixed,
\cref{eq:ndf} reduces to the Beckmann NDF with roughness
$\alpha=\sqrt{2\epsilon_h}$ on the upper hemisphere
(\cref{app:beckmannlimit}):
\begin{equation}
  \NDF[\SurfNorm]
  \xrightarrow{\epsilon_z\to0}
  D_{\mathrm{Beckmann}}(\SurfNorm)
  \coloneq
  \frac{\euler^{-\tan^2\theta/\alpha^2}}
       {\pi\alpha^2\cos^4\theta}.
  \label{eqn:beckmann_limit}
\end{equation}
The density vanishes on the lower hemisphere.
The one-sided projected area furthermore recovers the
Smith masking function (\cref{app:beckmannsmith}):
\begin{equation}
  \ProjAreaMinus[-\OutDir]
  \xrightarrow{\epsilon_z\to0}
  \frac{\cos\theta_\ast}
       {G_1^{\mathrm{Smith}}(\theta_\ast)}.
  \label{eqn:brdf_beckmann_projected_area_limit}
\end{equation}
Substituting these limits into \cref{eqn:vndf} gives the
Beckmann--Smith visible NDF \citep{Heitz:2014:Understanding}.
The GPIS transmittance also recovers Smith's
height-conditional masking function
(\cref{app:smith-height-conditional-masking}).
Consequently, \cref{eqn:brdf_single_full} reduces to the
classical Beckmann single-scattering BRDF with Smith's
joint masking-shadowing function, while extending
to BRDF-compatible GPISes beyond height fields.

\subsection{Extending to Multiple Scattering}
\label{sec:multiple-scattering}

Multiple scattering follows by tracing successive free flights
and scattering events, with each normal sampled
from the GPIS vNDF. The extinction factorization established
above lets us extend the random walk of
\citet{Heitz:2016:Multiplescattering} to our full-sphere NDFs.

To establish reciprocity of the resulting BRDF, we compare
each path with its reverse. Reversing the ray direction at
fixed height $z$ gives
$\ExtCoeff[z,-\RayDir]-\ExtCoeff[z,\RayDir]
=\rho^{\mathrm{Smith}}(z)(\RayDir\cdot\hat{\mathbf z})$
by \cref{eqn:asymmetry}.
Using $\dif z=(\RayDir\cdot\hat{\mathbf z})\,\dif t$
and $\rho^{\mathrm{Smith}}(z)\,\dif z=\dif u/u$ gives
\begin{equation}
  \frac{\Trans[\mathbf a\to\mathbf b]}
       {\Trans[\mathbf b\to\mathbf a]}
  =
  \exp\!\left[
    \int_{z_a}^{z_b}\rho^{\mathrm{Smith}}(z)\,\dif z
  \right]
  =
  \frac{u(z_b)}{u(z_a)},
\end{equation}
where $z_a$ and $z_b$ are the endpoint heights.
Along a complete path, the interior $u$ factors cancel,
and the endpoint factors agree on the common boundary.
For a reciprocal micro-BRDF, \cref{eqn:weighted_reciprocity}
then establishes BRDF reciprocity at every scattering order.

The occupancy factor also simplifies sampling.
Following \citet{Dupuy:2016:Additional}, we absorb it
through a change of coordinates, obtaining an equivalent
homogeneous half-space formulation. We develop this
remapping and the resulting sampling procedure in
\cref{sec:multiple-scattering-walk}.

\subsection{Full-Sphere Beckmann and GGX Families}
\label{sec:fullsphere-ndfs}

We call the Gaussian-gradient NDF of \cref{eq:ndf} the
\emph{full-sphere Beckmann} family. Applying the exponential
precision mixture of \cref{sec:mixture} yields the corresponding
\emph{full-sphere GGX} family. Both recover their classical
counterparts in the height-field limit.

Under the multiple-scattering walk above, these families yield
energy-preserving, reciprocal BRDFs for a nonabsorbing,
reciprocal micro-BRDF. They also support both in-plane and
out-of-plane anisotropy. The latter is important for capturing
the appearance of materials such as wood
\citep{Marschner:2005:Measuring}.

Previous work has explored full-sphere NDFs to avoid the
increasingly spiky geometry of height-field models at high
roughness and to connect rough surfaces with porous media
\citep{Dupuy:2016:Additional}. These include models based on
von Mises--Fisher distributions
\citep{dEon:2016:Anisotropic,dEon:2024:VMF}, whose projected
areas have been represented using series expansions and
approximations. Angular null scattering provides an unbiased
sampling approach for general full-sphere NDFs
\citep{dEon:2023:StudentT}, but can be costly.

Our families have closed-form projected areas
(\cref{eqn:one_sided_area_closed,eqn:mixture_projarea}) and
exact vNDF samplers developed in the next section.
\Cref{fig:high_roughness,fig:brdf_tilted} illustrate the
resulting multiply scattered appearance at high roughness
and with in-plane and out-of-plane anisotropy.

\section{Monte Carlo Rendering}
\label{sec:rendering}

We render GPISes by Monte Carlo integration of the anisotropic
RTE, using the transport coefficients derived in \cref{sec:forward}.
This requires sampling free-flight distances and scattering
directions, with the latter sampled through the vNDF and micro-BRDF.

\subsection{Free-Flight Sampling}
\label{sec:freeflight}

\paragraph{Homogeneous media}
With spatially constant GP statistics, extinction is constant
along each ray, so free-flight distances follow an exponential
distribution with rate $\ExtCoeff$.

\paragraph{Heterogeneous media}
When the GP statistics vary spatially, $\ExtCoeff$ varies along
the ray and free-flight sampling uses standard techniques for
participating media~\cite{Novak:2018:Monte}: ray marching, whose
bias is controlled by the step size, or null-collision (delta)
tracking, which avoids that bias.
Each candidate point requires one closed-form evaluation of
\cref{eqn:extinction_coeff} from the local statistics
$(\MeanF, \StdF, \CondGradMean, \CondGradCov)$. 
Both the occupancy factor and the directional factor
$\ProjAreaMinus[\mathbf{r}(t),\RayDir]$ are re-evaluated
because they can vary spatially.

Null-collision methods require a majorant valid over both
position and ray direction.
Although the occupancy factor is unbounded as
$\MeanF/\StdF\to-\infty$, its growth is only asymptotically
linear in $-\MeanF/\StdF$ for fixed $\StdF$.
On bounded rendering domains where $\MeanF/\StdF$ has a
finite minimum and $\StdF$ is bounded away from zero,
the occupancy factor remains bounded.
Together with bounded conditional gradient statistics,
this permits finite extinction majorants.
We precompute per-cell majorants via dense spatio-directional
sampling and store them in a coarse majorant grid.
Tighter, direction-dependent majorants could reduce
null-collision overhead.

Transmittance estimates, needed for next-event estimation,
can similarly use existing techniques.

\subsection{Sampling the Phase Function}
\label{sec:phasefnsampling}

Following the standard microflake construction
\citep{Heitz:2015:SGGX,Jakob:2010:Radiative}, we sample the
phase function in \cref{eqn:phasefn} by first drawing a normal
$\SurfNorm$ from the vNDF $\VNDF[\RayDir]$, then sampling
$\ScatDir$ using the micro-BRDF
$f_r(-\RayDir,\ScatDir;\SurfNorm)$.
For a perfect mirror micro-BRDF, reflection is deterministic
given $\SurfNorm$, with unit sample weight.

\paragraph{Gradient-space sampling}
The vNDF is the distribution of the normalized gradient at a
down-crossing (\cref{sec:ndfderivation}). We can therefore
sample an unnormalized gradient
$\GradField\in\mathbb{R}^3$ from the Kac--Rice-weighted Gaussian
\begin{align}
  \label{eqn:qg}
  q(\mathbf{g}) \;\propto\;
    \langle-\RayDir,\mathbf{g}\rangle\,\mathcal{N}\!\left(\mathbf{g};\,\CondGradMean,\,\CondGradCov\right)
\end{align}
and normalize it ($\SurfNorm=\GradField/\|\GradField\|$) to exactly sample from $\VNDF[\RayDir]$.

\paragraph{Factorization and algorithm}
The crossing weight depends only on the gradient component
along $\RayDir$. We decompose
$\GradField=\CondDeriv\RayDir+\GradPerp$, where
$\CondDeriv=\GradField\cdot\RayDir$ and $\GradPerp$ lies in the
plane perpendicular to $\RayDir$.
The scalar marginal is weighted by $|\CondDeriv|$ and restricted
to $\CondDeriv<0$, while the conditional distribution of
$\GradPerp$ given $\CondDeriv$ remains Gaussian.
The crossing-weighted distribution therefore factorizes as
\begin{align}
  \label{eqn:factorization}
  q(\CondDeriv, \GradPerp)
    \;\propto\;
    \underbrace{|\CondDeriv|\,\mathcal{N}(\CondDeriv;\,\CondDerivMean,\,\CondDerivVar)\,\mathbf{1}[\CondDeriv<0]}_{q_z(\CondDeriv)}
    \;\cdot\;
    \underbrace{\mathcal{N}_2\!\left(\GradPerp;\,\CondMeanPerp,\,\CondCovPerp\right)}_{q_{\perp|z}(\GradPerp)}.
\end{align}
Here, $\CondMeanPerp$ and $\CondCovPerp$ are the conditional
mean and covariance of $\GradPerp$ given $\CondDeriv$.
Algorithm~\ref{alg:samplevndf} first samples $\CondDeriv$ from
$q_z$, then draws $\GradPerp$ from the conditional Gaussian
$q_{\perp|z}$ and normalizes the reconstructed gradient.
The weighted 1D marginal $q_z$ requires the dedicated sampler
we develop next.

\begin{algorithm}
\caption{$\mathrm{SampleVNDF}(\RayDir,\,\CondGradMean,\,\CondGradCov,\,\xi_1,\,\xi_2,\,\xi_3)$}
\label{alg:samplevndf}
\begin{algorithmic}[1]
\State $\CondDerivMean \gets \RayDir^\mathsf{T}\CondGradMean$;\quad
       $\CondDerivStd  \gets \sqrt{\RayDir^\mathsf{T}\CondGradCov\RayDir}$
\State $\CondDeriv \gets \mathrm{SampleProjectedDerivative}(\CondDerivMean,\,\CondDerivStd,\,\xi_1)$
\State $\CondMeanPerp, \CondCovPerp \gets$ Gaussian conditioning on $\CondDeriv$
\State $\GradPerp \gets \mathrm{SampleGaussian2D}(\CondMeanPerp,\,\CondCovPerp,\,\xi_2,\,\xi_3)$
\State \Return $\SurfNorm \gets (\CondDeriv\RayDir + \GradPerp)\;/\;\|\CondDeriv\RayDir + \GradPerp\|$
\end{algorithmic}
\end{algorithm}

\paragraph{Sampling $q_z$}
Setting $u = -\CondDeriv > 0$, the marginal $q_z$ becomes
$q_u(u) \propto u\,\mathcal{N}(u;\,-\CondDerivMean,\,\CondDerivVar)$ for $u > 0$
--- a linearly-weighted truncated Gaussian.
Its CDF $F(U)$ is
\begin{align}
  \label{eqn:qz_cdf}
  \!\!\!F(U) &= \frac{G(U)}{\ProjAreaMinus[\RayDir]},\\
  \!\!\!G(U) &=
    \CondDerivStd\!\left[\normPDF[\tfrac{\CondDerivMean}{\CondDerivStd}]
      - \normPDF[\tfrac{U+\CondDerivMean}{\CondDerivStd}]\right]
    - \CondDerivMean\!\left[\normCDF[\tfrac{U+\CondDerivMean}{\CondDerivStd}]
      - \normCDF[\tfrac{\CondDerivMean}{\CondDerivStd}]\right],\!\!
\end{align}
where $\ProjAreaMinus[\RayDir]$ is the normalization already derived in \cref{sec:kacrice}.
Although this CDF has no closed-form inverse, we can sample it in $O(1)$ expected cost via the three-regime
rejection sampler in Algorithm~\ref{alg:sampleqz}. 

\begin{algorithm}
\caption{$\mathrm{SampleProjectedDerivative}(\CondDerivMean,\,\CondDerivStd,\,\xi)$}
\label{alg:sampleqz}
\begin{algorithmic}[1]
\State $\mu^* \gets -\CondDerivMean$;\quad $a \gets \mu^*/\CondDerivStd$
\If{$|a| < \varepsilon$} \Comment{Case 1: $\mu^*\approx 0$---Rayleigh}
  \State $u \gets \CondDerivStd\sqrt{-2\ln\xi}$
\ElsIf{$a > 0$} \Comment{Case 2: $\mu^*>0$---Gaussian proposal}
  \State $\tau \gets \sqrt{2}$;\quad $\lambda \gets \tfrac{1}{2}$;\quad
         $r^* \gets \tfrac{1}{2}(a + \sqrt{a^2+8})$;\quad $x^* \gets r^* - a$
  \Repeat
    \State Draw $r \gets a + \tau\,\mathcal{N}(0,1)$ \Comment{$\mathcal{N}(0,1)$ via Box-Muller}
    \State $\mathit{accept} \gets (r/r^*)\exp\!\left(-\tfrac{\lambda}{2}((r-a)^2-(x^*)^2)\right)$
  \Until{$r > 0$ \textbf{and} $\mathcal{U}(0,1) < \mathit{accept}$}
  \State $u \gets \CondDerivStd \cdot r$
\ElsIf{$|a| \leq 1$} \Comment{Case 3a: $0 < b{=}|a|\leq 1$---Rayleigh proposal}
  \Repeat
    \State Draw $r \gets \sqrt{-2\ln\mathcal{U}(0,1)}$
  \Until{$\mathcal{U}(0,1) < \exp(-b\,r)$}
  \State $u \gets \CondDerivStd \cdot r$
\Else \Comment{Case 3b: $b{=}|a|>1$---Gamma$(2,\,1/b)$ proposal}
  \Repeat
    \State Draw $r \gets -(\ln\mathcal{U}(0,1) + \ln\mathcal{U}(0,1))/b$
  \Until{$\mathcal{U}(0,1) < \exp(-r^2/2)$}
  \State $u \gets \CondDerivStd \cdot r$
\EndIf
\State \Return $\CondDeriv \gets -u$
\end{algorithmic}
\end{algorithm}
The proposal adapts to whether the conditional gradient
favors down-crossings or places them in the Gaussian tail.

\paragraph{Exact sampling for the heavy-tailed NDF}
The same decomposition into ray-parallel and perpendicular
components applies to the full-sphere GGX NDF in
\cref{eqn:mixture_ndf}.
Remarkably, its crossing-weighted scalar marginal has a CDF that can be
inverted in closed form by solving a quadratic
(\cref{app:mixture_sampling}).
Conditioned on $\CondDeriv$, the perpendicular distribution
is heavy-tailed, and its sampling procedure reuses the
Gaussian-conditioning parameters $\CondMeanPerp$ and
$\CondCovPerp$.
Algorithm~\ref{alg:sampleqz_ggx} combines these steps to
sample the vNDF exactly using three uniform random
variables and no rejection.

\begin{algorithm}
\caption{$\mathrm{SampleGGXVNDF}(\RayDir,\,\CondGradMean,\,\CondGradCov,\,\xi_1,\,\xi_2,\,\xi_3)$}
\label{alg:sampleqz_ggx}
\begin{algorithmic}[1]
\State $\CondDerivMean \gets \RayDir^\mathsf{T}\CondGradMean$;\quad
       $\CondDerivVar \gets \RayDir^\mathsf{T}\CondGradCov\RayDir$;\quad
       $\alpha_{\mathrm{eff}}^2 \gets 2\CondDerivVar$
\State $c \gets -\CondDerivMean$;\quad $R \gets \sqrt{c^2+\alpha_{\mathrm{eff}}^2}$
\State $h \gets \xi_1(c+R) - R$
\State $s \gets \sqrt{\max(0,\,R^2-h^2)}$
\State $q \gets \alpha_{\mathrm{eff}}\,
  \dfrac{\alpha_{\mathrm{eff}}\,s+c\,h}
        {c\,s-\alpha_{\mathrm{eff}}\,h}$
\State $\CondDeriv \gets -(c+q)$
\State $\CondMeanPerp$, $\CondCovPerp \gets$ Gaussian conditioning on $\CondDeriv$
\State $\beta \gets 1 + q^2/(2\CondDerivVar)$
       \Comment{$= 1+(\CondDeriv-\CondDerivMean)^2/(2\CondDerivVar)$}
\State $\rho \gets \sqrt{2\beta\bigl[(1-\xi_2)^{-2/3}-1\bigr]}$;\quad $\varphi \gets 2\pi\xi_3$
\State $\boldsymbol L \gets \mathrm{Cholesky}(\CondCovPerp)$
\State $\GradPerp \gets \CondMeanPerp + \boldsymbol L\,\rho\,(\cos\varphi,\,\sin\varphi)^\mathsf{T}$
\State \Return $\SurfNorm \gets (\CondDeriv\RayDir + \GradPerp)\;/\;\|\CondDeriv\RayDir + \GradPerp\|$
\end{algorithmic}
\end{algorithm}

\paragraph{Connection to prior work on GPIS rendering}
The factorization in \cref{eqn:factorization} parallels the
gradient decomposition used by \citet{Xu:2025:Practical} and
\citet{Shi:2026:Conditional}.
Their realization-based renderers obtain the directional
derivative at an intersection by ray marching, then sample
the perpendicular gradient components from a conditional
Gaussian.
Under the local-conditioning approximation, we instead draw
$\CondDeriv$ directly from the weighted marginal $q_z$,
then sample $\GradPerp$ from $q_{\perp|z}$.
This requires no explicit realization, but replaces the
first-hit statistics of ray marching with the typical-crossing
law of \cref{sec:ndfderivation}.

\citet{Huang:2026:Macrofacet} use a mixture of Beckmann-vNDF
and uniform-hemisphere proposals to sample their vNDF.
Our gradient-space factorization enables exact sampling
of the GPIS vNDF.

\begin{figure}[t]
  \centering
  \setlength{\tabcolsep}{1pt}
  \renewcommand{\arraystretch}{0.5}
  \begin{tabular}{@{}r@{\hspace{1pt}}cccc@{}}
    \rotatebox{90}{\makebox[0.23\linewidth][c]{\footnotesize Gaussian}} &
    \includegraphics[width=0.23\linewidth]{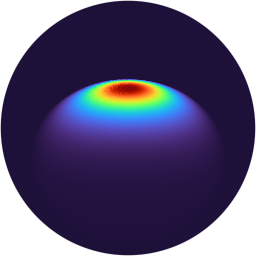} &
    \includegraphics[width=0.23\linewidth]{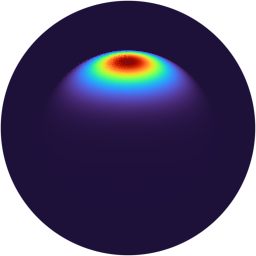} &
    \includegraphics[width=0.23\linewidth]{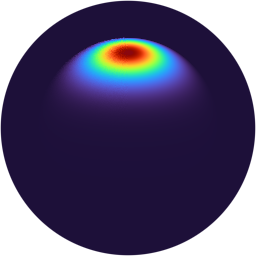} &
    \includegraphics[width=0.23\linewidth]{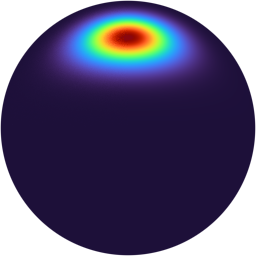} \\
    \rotatebox{90}{\makebox[0.23\linewidth][c]{\footnotesize Heavy-tailed}} &
    \includegraphics[width=0.23\linewidth]{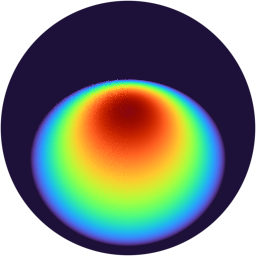} &
    \includegraphics[width=0.23\linewidth]{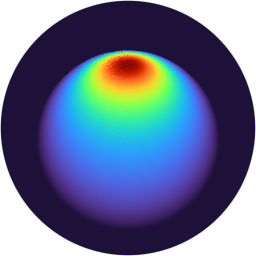} &
    \includegraphics[width=0.23\linewidth]{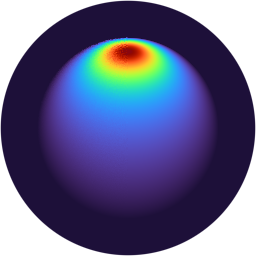} &
    \includegraphics[width=0.23\linewidth]{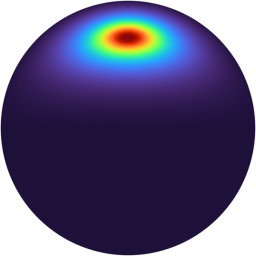} \\[2pt]
     &
    \makebox[0.23\linewidth][c]{\footnotesize $\theta = -45^\circ$} &
    \makebox[0.23\linewidth][c]{\footnotesize $\theta = -15^\circ$} &
    \makebox[0.23\linewidth][c]{\footnotesize $\theta = 0^\circ$} &
    \makebox[0.23\linewidth][c]{\footnotesize $\theta = 90^\circ$} \\
  \end{tabular}
  \caption{Validation of the exact vNDF importance samplers: Monte Carlo estimates from each sampler (left halves) match the corresponding closed-form vNDFs (right halves). Top: the sampler for the Gaussian gradient distribution (Algorithm~\ref{alg:samplevndf}, using Algorithm~\ref{alg:sampleqz}); the four incidence angles exercise all four proposal branches of Algorithm~\ref{alg:sampleqz}. Bottom: heavy-tailed sampler (Algorithm~\ref{alg:sampleqz_ggx}) against the heavy-tailed vNDF \eqref{eqn:mixture_ndf}. All panels use $(\alpha,\ell_z,\partial\mu/\partial z)=(1,1,2)$; $\theta=90^\circ$ is normal incidence and $\theta=0^\circ$ grazing.}
  \label{fig:vndf-is-validation}
\end{figure}

\paragraph{Sampler validation}
Both samplers agree with their analytic vNDFs in
\cref{fig:vndf-is-validation}.
The asymmetric configuration and incidence sweep exercise
all four proposal branches of Algorithm~\ref{alg:sampleqz}
and all sign regimes of the heavy-tailed CDF inversion.

\subsection{Multiple-Scattering Surface Walk}
\label{sec:multiple-scattering-walk}

We evaluate the single-scattering BRDF analytically using
\cref{eqn:brdf_single_full} and importance sample it with the
vNDF samplers of \cref{sec:phasefnsampling}.
For multiple scattering, the extinction factorization in
\cref{sec:single-scattering-brdf} has the Smith microflake form
used by \citet{Heitz:2016:Multiplescattering}, with
$\ProjAreaMinus[\RayDir]$ now obtained from the broader class
of BRDF-compatible GPISes.

Following previous work \citep{Dupuy:2016:Additional,Lumme:1990:Diffuse}, we absorb the
occupancy factor by remapping height to
$z_{\mathrm h}\coloneq\log u=\log\normCDF[z/\StdF]$,
where $u$ is the empty-space probability coordinate introduced
in \cref{sec:single-scattering-brdf}.
Since
$\dif z_{\mathrm h}=\dif u/u=\rho^{\mathrm{Smith}}(z)\,\dif z$,
the remapping yields a homogeneous medium with
direction-dependent extinction $\ProjAreaMinus[\RayDir]$.
The range $z\in\mathbb{R}$ maps to
$z_{\mathrm h}\in(-\infty,0)$, with $z_{\mathrm h}=0$
corresponding to the escape boundary.
The walk operates entirely in $z_{\mathrm h}$ without
recovering physical height.
Although $\StdF$ controls the physical thickness of the
transition region, it does not appear in the remapped walk.

We apply the random walk of
\citet{Heitz:2016:Multiplescattering} in the remapped coordinates
(Algorithm~\ref{alg:samplegpiswalk}).
For incident direction $\OutDir[i]$, the walk starts at
$z_{\mathrm h,0}=0$ with $\RayDir_1=-\OutDir[i]$.
Each step samples an exponential free flight with rate
$\ProjAreaMinus[\RayDir]$ for the current direction.
If the flight reaches the escape boundary $z_{\mathrm h}=0$,
we return the outgoing direction and accumulated throughput.
Otherwise, we sample a vNDF normal, scatter according to the
micro-BRDF, update the throughput, and repeat.
Free-flight sampling accounts for masking and shadowing
without an explicit height-conditional masking function.

\begin{algorithm}
\caption{$\mathrm{SampleGPISBRDF}(\OutDir[i],\,\NDF,\,f_r)$}
\label{alg:samplegpiswalk}
\begin{algorithmic}[1]
\State $\RayDir_1\gets-\OutDir[i]$;\quad
       $z_{\mathrm h,0}\gets0$;\quad
       $\mathrm{weight}\gets1$;\quad $r\gets1$
\While{true}
  \State Draw $t_r\sim
    \mathrm{Exp}\!\left(\ProjAreaMinus[\RayDir_r]\right)$
    \Comment{Homogeneous free flight}
  \State $z_{\mathrm h,r}\gets z_{\mathrm h,r-1}
    +t_r(\RayDir_r\cdot\hat{\mathbf z})$
  \If{$z_{\mathrm h,r}\geq0$}
    \State \Return $(\RayDir_r,\mathrm{weight})$
      \Comment{Direction and weight}
  \EndIf
  \State Draw $(\RayDir_{r+1},w_{r+1})\gets
    \mathrm{SamplePhase}(\RayDir_r,\NDF,f_r)$
  \State $\mathrm{weight}\gets w_{r+1}\,\mathrm{weight}$;\quad
         $r\gets r+1$
\EndWhile
\end{algorithmic}
\end{algorithm}

Here, $\mathrm{weight}$ is the accumulated path throughput.
$\mathrm{SamplePhase}$ samples a vNDF normal and the
micro-BRDF, returning the scattered direction and the
importance weight $w_{r+1}$ for that collision, including
absorption.  A position-free evaluation of the GPIS BRDFs can reduce variance
in some cases using the approach of \citet{Bitterli:2022:Positionfree}.  The
related formulation by \citet{Cui:2023:Multiplebounce}, however, assumes an NDF
restricted to the hemisphere and a specular microsurface, so does not generally apply.

\section{Accuracy of the Local-Conditioning Approximation}
\label{sec:lca_accuracy}

The local-conditioning approximation conditions on the current
point being in empty space, but discards how earlier observations
along the ray alter the field distribution ahead.
Its accuracy therefore depends on how quickly this spatial memory
fades relative to the distance between crossings.

\subsection{Memory Models}
\label{sec:memory}

\citet{Seyb:2024:Microfacets} introduced a hierarchy of GPIS transport
models that trade memory, and therefore computational cost, for
accuracy.
All sample a correlated field along the current ray segment,
but differ in the information retained between segments:
Global retains field values from all previous segments,
Renewal retains only $\GPField=0$ at the most recent collision,
and Renewal+ additionally retains the gradient there.

The LCA instead replaces field sampling along the ray with
a local crossing rate conditioned only on the current point
being in empty space.
This discards conditioning on earlier observations both
within and between ray segments, yielding local transport
coefficients compatible with the classical RTE.
Retaining such observations would require additional
transport state, as in the
GLBE~\citep{Larsen:2011:Generalized,Bitterli:2018:Radiative}.

\begin{figure}
  \centering
  \includegraphics[width=\linewidth]{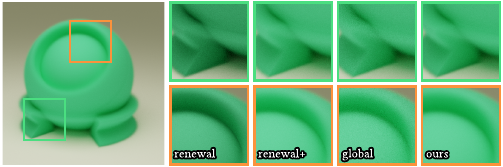}
  \caption{The same scene rendered with different memory models. The full render (left) uses our approach. The models are visually similar overall, with the clearest differences appearing in concave regions where recent ray history contains useful information about nearby geometry.}
  \label{fig:memory_model_comp}
\end{figure}
\begin{figure*}[t]
  \centering
  \includegraphics[width=\linewidth]{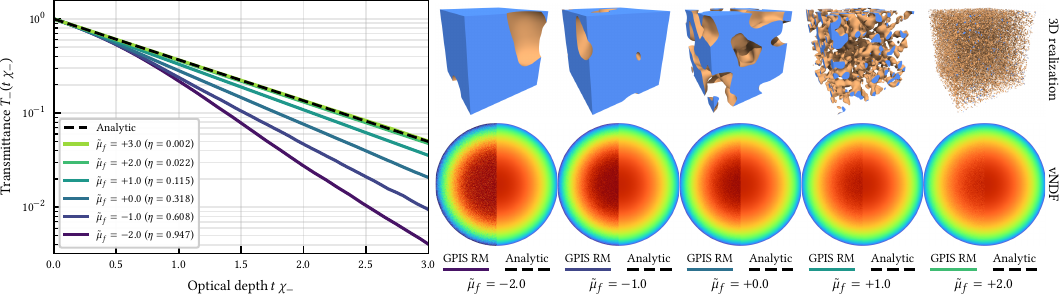}
  \caption{Local-conditioning approximation (LCA) accuracy across void levels $\VoidLevel$ for a homogeneous GPIS medium. \emph{Left:} transmittance $\Trans(t\,\ExtCoeff)$ from first-crossing Monte Carlo sampling (solid) against the analytic LCA prediction $\exp(-t\,\ExtCoeff)$ (dashed). Legend entries report the correlation-to-chord ratio $\eta=\CorLen\ExtCoeff$ from \cref{eqn:mfp_ratio}. \emph{Right:} a single realization (top) and the corresponding vNDF (bottom) for each void level. The cubical crops have an optical depth of 3 in each dimension. Each vNDF is split between the empirical ensemble estimate from ray marching (left) and the analytic LCA prediction (right).}
  \label{fig:void-level-sweep-combined}
\end{figure*}

\Cref{fig:memory_model_comp} compares these memory models.
Renewal can resample a surface whose solid side contains the
outgoing ray, causing immediate re-intersections and visible
darkening.
Renewal+ uses the retained gradient to preserve the local surface
orientation, while the LCA avoids this through
its empty-space conditioning.
The remaining differences are most visible in crevices, where
correlations with nearby geometry matter.
The next subsection quantifies when this discarded spatial
memory affects accuracy.

\subsection{The Void Level and When Ray History Matters}
\label{sec:lca_crossings}

\paragraph{Void level} We begin with a dimensionless measure of how strongly the GPIS favors
empty space at a point. We define the \emph{void level} as the field mean
measured in units of its standard deviation,
\begin{align}
  \label{eqn:voidlevel_def}
  \VoidLevel[\mathbf{x}]
  \;\coloneq\;
  \frac{\GPMean[\mathbf{x}]}{\StdF[\mathbf{x}]},
\end{align}
and the corresponding \emph{void fraction} as
\begin{align}
  \label{eqn:voidfrac_def}
  \VoidFrac[\mathbf{x}]
  \;\coloneq\;
  \normCDF[\VoidLevel[\mathbf{x}]]
  =
  \mathbb{P}\!\left(\GPField[\mathbf{x}]>0\right).
\end{align}
At $\VoidLevel=0$, empty and solid space are equally likely. Increasing
the void level makes the empty phase more probable, moving the GPIS toward a more dilute regime.

\paragraph{Void chords and ray history}
Along a ray, successive zero crossings mark alternating empty and
solid intervals. We call a complete empty interval a \emph{void chord}
and denote its length by $C_{\mathrm v}$. For a stationary GPIS, the
relation between void fraction and chord length follows from simple
length accounting: along a long ray, the total length in empty space is
the number of void chords times their average length. Dividing by the
ray length turns these quantities into the void fraction, down-crossing
density, and mean chord length, respectively
\citep[Sec.~II]{Roberts:1999:Chorddistribution}:
\begin{align}
  \label{eqn:mean_void_chord}
  \VoidFrac
  = \DownCross\,\mathbb{E}[C_{\mathrm v}]
  \qquad\Longrightarrow\qquad
  \mathbb{E}[C_{\mathrm v}]
  = \frac{\VoidFrac}{\DownCross}
  = \frac{1}{\ExtCoeff},
\end{align}
The final equality follows from the LCA definition
$\ExtCoeff=\DownCross/\VoidFrac$ in \cref{eqn:local_conditioning}.
Under stationarity, the LCA extinction is constant along a
ray, giving exponential free flights.
The exact GPIS crossing rate, however, depends on the ray's
survival history, so its free-flight distribution is generally
non-exponential.
Although the mean LCA free flight equals the mean length of
complete void chords, it need not equal the mean remaining
distance to a crossing from a point sampled in empty space.

Whether this replacement is accurate depends on how the mean void-chord
length compares with the distance over which the GP retains memory. Let
$\CorLen$ denote the kernel correlation length, and define the
dimensionless \emph{correlation-to-chord ratio}
\begin{align}
  \label{eqn:memory_ratio}
  \eta
  \;\coloneq\;
  \frac{\CorLen}{\mathbb{E}[C_{\mathrm v}]}
  =
  \CorLen\ExtCoeff.
\end{align}
Equivalently, $1/\eta=\mathbb{E}[C_{\mathrm v}]/\CorLen$ is the
mean void-chord length measured in correlation lengths.
When $\eta\ll1$, the mean void chord spans many correlation
lengths. For rapidly decaying kernels such as SE, this
separation of scales allows spatial memory to fade before
the next crossing.
When $\eta$ is not small, the mean void-chord length is
comparable to or shorter than the correlation length, and
recent history may remain informative at the next crossing.

\paragraph{Dependence on the void level}
For a homogeneous, isotropic, stationary GPIS with an SE kernel, the
ratio $\eta$ depends only on the void level. In this setting,
$\CondGradMean=\mathbf{0}$,
$\ProjAreaMinus=\CondDerivStd/\sqrt{2\pi}$, and
$\CondDerivStd=\StdF/\CorLen$. Substituting these quantities into
\cref{eqn:extinction_coeff} gives
\begin{align}
  \label{eqn:mfp_ratio}
  \eta
  = \CorLen\ExtCoeff
  = \frac{\normPDF[\VoidLevel]}
         {\sqrt{2\pi}\,\normCDF[\VoidLevel]}.
\end{align}
This ratio decreases monotonically with $\VoidLevel$. At
$\VoidLevel=0$, $\eta=1/\pi$, so the mean void chord spans
$\pi$ correlation lengths. At $\VoidLevel=1$ and $2$, it spans
approximately $8.7$ and $45$ correlation lengths, respectively, and
the ratio vanishes as $\VoidLevel\to\infty$.

This ratio provides a qualitative guide to LCA accuracy,
but not an error bound: the exact free-flight distribution
also depends on the full chord statistics and the shape
of the covariance kernel.

\paragraph{RTE-equivalent GPISes}
Different GPISes can induce the same LCA medium; we call
them \emph{RTE-equivalent}.
In the homogeneous, isotropic setting above, we construct
such a family by varying $\VoidLevel$ and setting
$\CorLen=\eta(\VoidLevel)/\ExtCoeff$ for a fixed $\ExtCoeff$,
using \cref{eqn:mfp_ratio}.
This preserves the extinction and phase function while
changing the GPIS correlation length, chord statistics,
and dependence on ray history.
We compare members of this family in \cref{sec:lca_validation}
to test how LCA accuracy varies with $\eta$;
\cref{sec:inversion} develops the general construction.

\begin{figure}[t]
  \centering
  \includegraphics[width=\linewidth]{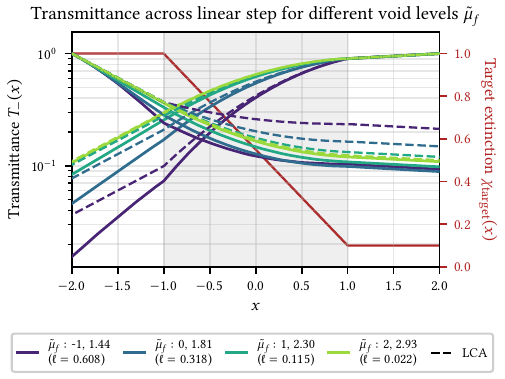}
  \caption{Transmittance through heterogeneous GPISes constructed from the
    target extinction profile shown in red (right axis). Solid curves are
    ensemble averages over explicit GPIS realizations, while dashed curves
    are the corresponding full, direction-dependent LCA predictions. Each
    color denotes a void-level configuration reported in the
    legend. Both traversal directions are shown, beginning at $\Trans=1$ on
    opposite sides. Solid and dashed curves nearly coincide at high void
    levels, while visible discrepancies remain at lower void levels.}
  \label{fig:void-level-sweep-linearstep}
\end{figure}

\subsection{Validation Against Realization Ensembles}
\label{sec:lca_validation}

\Cref{fig:void-level-sweep-combined} compares realization-based
transmittance and first-crossing vNDFs with the LCA
for homogeneous, RTE-equivalent GPISes.
Across this family, increasing $\VoidLevel$ produces finer,
more dilute realizations, while the LCA extinction and
phase function remain fixed.

At the ray origin, the transmittances share the same initial
slope because both condition only on the point being in
empty space. Further along the ray, survival provides
additional information that the LCA discards.
The realization-based transmittances lie below the LCA
prediction, with the largest discrepancy at $\VoidLevel=-2$,
where $\eta\approx0.95$.
As the void level increases and $\eta$ decreases, both the
transmittance and first-crossing vNDF approach their LCA
counterparts.

\paragraph{Long-range decay}
The nearly straight tails in the semilogarithmic plot of
\cref{fig:void-level-sweep-combined} are consistent with
asymptotically exponential persistence decay for stationary
SE fields~\citep[Theorem~1]{Feldheim:2025:Persistence}.
Intuitively, a long empty ray must remain in the void phase
across many weakly correlated regions, making such paths
exponentially rare.
The decay rate, or \emph{persistence exponent}, depends on
the full correlation structure of the field
\citep[Sec.~6.3]{Bray:2013:Persistence}, whereas the LCA rate
$\ExtCoeff$ uses only local crossing statistics.
An exponential tail therefore need not have the LCA decay
rate, as illustrated by the faster decay of the
realization-based transmittances.

\Cref{fig:void-level-sweep-linearstep} extends the comparison
to heterogeneous GPISes.
For each global correlation length, we use the homogeneous
relation in \cref{eqn:mfp_ratio} to convert a prescribed
linear extinction ramp into a spatially varying void level.
Within the ramp, the induced mean gradient changes the
projected area, so the full LCA extinction depends on both
the GPIS configuration and ray direction.
Each realization-based transmittance is therefore compared
with its own LCA prediction.
At high void levels, the curves nearly coincide in both
directions; discrepancies remain at lower void levels.
These results suggest that the correlation-to-chord ratio
remains a useful qualitative guide to LCA accuracy under
spatial variation.

More detailed Gaussian first-passage methods
\citep{Roberts:1999:Chorddistribution,Lindgren:2019:Gaussian}
and rough-surface shadowing models that account for spatial
correlations \citep{Kapp:1994:Effect} could better reproduce
realization-based transmittance.
Incorporating their history dependence into transport,
however, would require additional state beyond the local
coefficients of the classical RTE.

\section{From Participating Media to GPISes}
\label{sec:inversion}

\cref{sec:forward} derived the extinction and scattering laws induced by a
given GPIS under the LCA. We now consider the inverse problem: given a
target RTE medium, what GP mean $\GPMean[\mathbf{x}]$ and covariance
$\CovFn[\mathbf{x}][\mathbf{x}']$ reproduce its local transport
coefficients?

\subsection{The Inversion Problem and Equivalence Families}
\label{sec:inversion_equiv}

Matching a target medium requires reproducing both its
free-flight law, determined by $\ExtCoeff$, and its scattering
law, described by $\PhaseFunc$.
These transport coefficients do not uniquely determine the
GP: different fields can describe the same GPIS, and
geometrically different GPISes can induce the same LCA medium.

\paragraph{Exact GPIS equivalence: amplitude scaling}
Replacing $\GPField$ with $\lambda\GPField$ for any smooth field
$\lambda(\mathbf{x}) > 0$ is an exact symmetry of the GPIS: the zero
set $\{\lambda\GPField = 0\} = \{\GPField = 0\}$ is identical, and on
it $\nabla(\lambda\GPField) = \lambda\nabla\GPField$, so the surface
normals are unchanged too.
The NDF $\NDF$ and the one-sided projected areas $\ProjAreaMinus$ and
$\ProjAreaPlus$ each scale by $\lambda$, but this factor cancels in the
combinations that define the RTE quantities. Thus,
$\ExtCoeff$, $\PhaseFunc$, and $\VNDF[\RayDir]$ remain exactly
unchanged.

\paragraph{Reparameterized extinction}
Since $\GPMean[\mathbf{x}]$ and $\StdF[\mathbf{x}]$ can only enter RTE
quantities through amplitude-invariant combinations, the void level of
\cref{eqn:voidlevel_def} is the natural parameter. Written in terms of
$\VoidLevel$, the extinction coefficient \cref{eqn:extinction_coeff}
reads
\begin{align}
  \label{eqn:extinction_occparam}
  \ExtCoeff[\mathbf{x}, \RayDir]
    = \frac{\normPDF[\VoidLevel[\mathbf{x}]]}
           {\StdF[\mathbf{x}]\,\normCDF[\VoidLevel[\mathbf{x}]]}
      \cdot \ProjAreaMinus[\mathbf{x}, \RayDir],
\end{align}
where $\ProjAreaMinus$ is given by
\cref{eqn:one_sided_area_closed}.
The field amplitude cancels between the two factors, leaving
extinction determined by the void level $\VoidLevel$ and
the normalized conditional gradient statistics
$\CondGradMean/\StdF$ and $\CondGradCov/\FieldVar$
for a given ray direction.

\paragraph{RTE-equivalent families}
Field rescaling changes the GP but not the GPIS. Under the LCA, the RTE
equivalence of \cref{sec:lca_crossings} is a strictly weaker relation:
GPISes with different geometry can induce the same medium.
This arises in two ways. First, after fixing $\StdF[\mathbf{x}]$, the LCA reads the GP only through the pointwise triple
$(\VoidLevel[\mathbf{x}],\,\CondGradMean[\mathbf{x}],\,\CondGradCov[\mathbf{x}])$,
so any change to the kernel's long-range correlation that
preserves this triple leaves every RTE quantity unchanged while altering
the underlying geometry. This gives kernel-shape equivalence families.
Second, distinct triples can result in the same extinction and phase
function, as in the void-level families of \cref{eqn:mfp_ratio}. The inversion 
therefore returns an equivalence
family rather than a single GPIS.

\paragraph{Kernel-shape families}
For a smooth stationary isotropic covariance
$\CovFn[\mathbf{x}][\mathbf{x}']
=\FieldVar\,k(r)$, where
$r=\|\mathbf{x}-\mathbf{x}'\|$ and $k(0)=1$,
the covariance identities of \cref{sec:fieldgradientstats}
give
\begin{equation}
  \CondGradCov=\CovGrad=\FieldVar\,\KernelConst\,\mathbf{I},
  \quad\text{where}\quad \KernelConst\coloneq-k''(0).
  \label{eqn:kernel_curvature_stats}
\end{equation}

With the mean field and micro-BRDF fixed, the kernel
influences the LCA RTE only through its one-point variance
$\FieldVar$ and its curvature at zero lag $\KernelConst$.
Kernels with \emph{dramatically different shapes} therefore
produce identical LCA medium parameters when these two
quantities match.

\begin{figure}
  \centering
  \includegraphics[width=\linewidth]{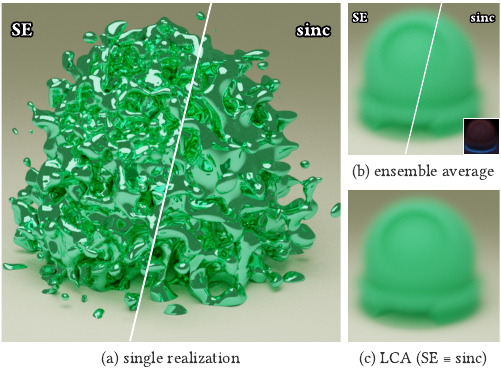}
  \caption{RTE-equivalent SE and sinc kernels with matched
variance and curvature $\KernelConst=-k''(0)$.
(a)~Single realizations with different microstructure.
(b)~Ensemble-average renders and their difference scaled by $10\times$
(red: SE brighter; blue: sinc brighter).
(c)~The identical LCA rendering for both kernels.}
  \label{fig:sinc_se_render}
\end{figure}

\Cref{fig:sinc_se_render} demonstrates this equivalence
with unit-variance SE and sinc kernels matched in
$\KernelConst$. Their realizations and ensemble-averaged
renderings differ, while their LCA renderings are identical.
We give the matching condition and realization procedure
in the supplemental material.

\begin{figure}
  \centering
  \includegraphics[width=\linewidth]{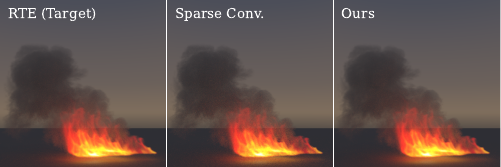}
  \caption{Converting an existing volumetric asset to a GPIS
using the inversion in \cref{sec:inversion_heterog}.
We compare the target RTE rendering (left),
realization-based GPIS rendering using sparse convolution
(center), and the corresponding LCA rendering (right).}
  \label{fig:vdb_to_gpis}
\end{figure}

\subsection{Homogeneous Media}
\label{sec:inversion_homog}

\paragraph{Isotropic media}
For a homogeneous target with constant extinction
$\ExtCoeff$ and isotropic scattering, we use a constant
mean and a stationary isotropic SE kernel.
These give $\CondGradMean=\mathbf{0}$ and
$\CondGradCov\propto\mathbf{I}$, yielding a uniform NDF
and, with a perfect mirror micro-BRDF, a uniform phase
function \eqref{eq:mirrorpf}.
The extinction match then reduces to
\begin{align}
  \label{eqn:corlen_iso}
  \frac{\normPDF[\VoidLevel]}{\normCDF[\VoidLevel]}
    = \sqrt{2\pi}\,\CorLen\,\ExtCoeff.
\end{align}
Since its left-hand side
decreases strictly from $+\infty$ to $0$, choosing either
$\VoidLevel$ or $\CorLen$ uniquely determines the other. The resulting
GPIS has mean $\GPMean=\VoidLevel\StdF$ and an isotropic SE covariance
with correlation length $\CorLen$, for any $\StdF>0$. 

\paragraph{SGGX media}
A homogeneous SGGX target has density
$\SGGXDensity>0$ and a symmetric positive-definite shape
matrix $\SGGXMat$.
As shown in \cref{app:sggxlimit},
$\CondGradMean=\mathbf{0}$ and
$\CondGradCov=2\pi\SGGXMat$ recover its NDF and projected
area.
Using the same micro-BRDF as the target then matches its
phase function.
Matching its extinction
$\ExtCoeff[\RayDir]=\SGGXDensity\ProjAreaMinus[\RayDir]$
using \cref{eqn:extinction_occparam} gives
\begin{align}
  \StdF
    = \frac{\normPDF[\VoidLevel]}
           {\SGGXDensity\,\normCDF[\VoidLevel]}.
  \label{eqn:sigf_sggx}
\end{align}
Choosing either $\VoidLevel$ or $\StdF$ uniquely determines the
other. The resulting GP has constant mean
$\GPMean=\StdF\VoidLevel$ and stationary anisotropic SE covariance.
\begin{align}
  \CovFn[\mathbf{x}][\mathbf{x}']
    = \FieldVar\,
      \exp\!\left(
        -\frac{\pi}{\FieldVar}
        (\mathbf{x}-\mathbf{x}')^{\mathsf{T}}
        \SGGXMat
        (\mathbf{x}-\mathbf{x}')
      \right).
  \label{eqn:gp_sggx_homog}
\end{align}
Because $\CondGradCov=2\pi\SGGXMat$ is fixed, varying
$\StdF$ also changes the kernel's correlation lengths.
\subsection{Heterogeneous Media}
\label{sec:inversion_heterog}

\paragraph{Isotropic targets}
We invert a scalar extinction field $\ExtCoeff[\mathbf{x}]$ through
a spatially varying mean and a stationary isotropic SE covariance
with fixed $\StdF$ and $\CorLen$. At each point, we solve
\cref{eqn:corlen_iso} for $\VoidLevel[\mathbf{x}]$ and set
$\GPMean[\mathbf{x}]=\StdF\VoidLevel[\mathbf{x}]$.
The extinction match becomes exact as $\CorLen\to0$, but is
approximate at finite correlation lengths because
\cref{eqn:corlen_iso} neglects the gradient of the void level.
We use this conversion for the heterogeneous targets in our results,
as illustrated in \cref{fig:vdb_to_gpis}.

\paragraph{Induced asymmetry}
Spatial variation in the void level induces a nonzero conditional
mean gradient, making the NDF asymmetric and introducing directional
dependence into the extinction. Using
$\StdF=\CorLen\CondDerivStd$, we can write
$\CondGradMean=\StdF\nabla\VoidLevel$ as
\begin{align}
  \label{eqn:isotropy_error}
  \frac{\|\CondGradMean[\mathbf{x}]\|}{\CondDerivStd}
    = \CorLen\|\nabla\VoidLevel[\mathbf{x}]\|.
\end{align}
When the void level changes little over one correlation length,
$\CondGradMean$ is small relative to the random gradient
fluctuations. The NDF is then nearly isotropic, and the extinction
closely matches the target.

\paragraph{Exact asymmetric-NDF constructions}
Although induced asymmetry is undesirable for isotropic targets,
it can be used to match asymmetric target NDFs. This extends our
inverse construction beyond classical symmetric microflake models
such as SGGX. 

Consider a target from the GPIS family of \cref{sec:ndfderivation},
with spatially varying $\CondGradMean[\mathbf{x}]$ and constant
$\CondGradCov$. We fix $\StdF>0$, choose a stationary covariance
matching $\CondGradCov$, and seek a void-level field satisfying
\begin{align}
  \nabla\VoidLevel[\mathbf{x}]
    = \frac{\CondGradMean[\mathbf{x}]}{\StdF}.
\end{align}
This requires $\CondGradMean/\StdF$ to be a gradient field,
whose scalar potential determines $\VoidLevel$ up to an additive
constant. Setting $\GPMean[\mathbf{x}]=\StdF\VoidLevel[\mathbf{x}]$
then matches the target NDF exactly.
Varying the additive constant yields a corresponding family
of extinction fields through \cref{eqn:extinction_occparam}.
A prescribed extinction field can be matched simultaneously
only if it belongs to this family.

For a spatially constant asymmetric NDF
($\CondGradMean=\mathbf{c}\neq\mathbf{0}$), the compatibility
condition holds automatically and the required mean is linear:
$\GPMean[\mathbf{x}]=\mathbf{c}\cdot\mathbf{x}+c_0$.

\paragraph{Constant-void-level alternative}
Alternatively, holding $\VoidLevel$ constant gives
$\CondGradMean=\mathbf{0}$, allowing pointwise matches to isotropic
and SGGX targets through spatially varying covariance statistics.
However, realizing these statistics with a globally valid
nonstationary covariance is nontrivial.
The probability of empty space is also uniform, and there is no
distinct mean surface.
We therefore do not use this construction in our results.

\subsection{Choosing Among a Family of GPISes}
\label{sec:inversion_approx}

For the isotropic constructions above, decreasing $\CorLen$
at fixed target extinction raises $\VoidLevel$ through
\cref{eqn:corlen_iso}. This changes the underlying geometry
while providing two benefits:
\begin{enumerate}
  \item \textbf{Improved LCA accuracy.}
        The correlation-to-chord ratio $\eta$ decreases
        \eqref{eqn:mfp_ratio}, moving the GPIS toward the dilute
        regime validated in \cref{sec:lca_validation}.
  \item \textbf{Reduced NDF asymmetry.}
        For the heterogeneous conversion, the residual
        $\CorLen\|\nabla\VoidLevel\|$ decreases
        \eqref{eqn:isotropy_error}, bringing the NDF and extinction
        closer to their isotropic targets.
\end{enumerate}

\begin{figure*}
  \centering
  \includegraphics[width=\textwidth]{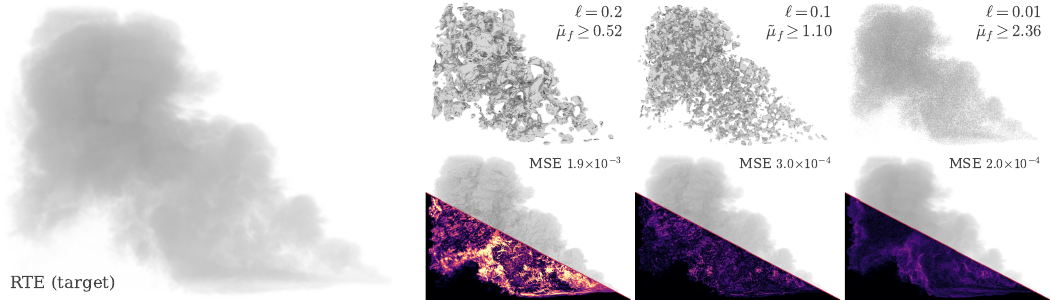}
  \caption{Choosing the void level for heterogeneous media.
We convert the target RTE medium (left) using successively
smaller global correlation lengths $\CorLen$, which raise
the void level.
Single GPIS realizations (top) change from dense,
interlocking structures to finer, sparser grains.
The corresponding LCA renderings (bottom) approach the
target, with decreasing MSE.
Per-pixel absolute luminance differences are shown on a
shared scale, with brighter values indicating larger error.
The increasingly dilute geometry also favors LCA accuracy
(\cref{sec:lca_validation}).}
  \label{fig:smoke_void_level_sweep}
\end{figure*}

\Cref{fig:smoke_void_level_sweep} compares three conversions
of the same heterogeneous smoke medium using successively
smaller global correlation lengths.
The realizations change from dense, interlocking structures
to finer, sparser grains.
As the induced NDF asymmetry decreases, the LCA render
approaches the target RTE render, with image MSE falling
by roughly an order of magnitude across the sweep.

We recommend choosing $\CorLen$ so that both
$\CorLen\ExtCoeff[\mathbf{x}]$ and
$\CorLen\|\nabla\VoidLevel[\mathbf{x}]\|$ are small in the regions
of interest. Shorter correlation lengths improve approximation
accuracy while producing finer, sparser geometry, so the final
choice also depends on the desired geometric structure.

\section{Implementation and Results}
\label{sec:results}

\subsection{Implementation}
We extend the GPIS implementations of
\citet{Seyb:2024:Microfacets} and \citet{Xu:2025:Practical},
built on Tungsten \citep{Bitterli:2018:Tungsten}.
Our method uses standard volume rendering, with extinction
and phase functions computed from pointwise GP statistics.
All equal-time experiments used delta tracking
and ran on an AMD Ryzen~9 5900XT CPU
(16 cores, 32 threads).

The supplemental material includes a browser-based GPIS
explorer for normal distributions and scattering.
It allows variation of GP parameters, comparing analytic and
Monte Carlo results alongside Beckmann, SGGX, and GGX
references, and inspecting the corresponding 3D single
realizations.

We used AI assistance in developing the supplemental
GPIS explorer, testing our Tungsten code, and checking
mathematical derivations. We developed the research ideas
and methods and reviewed and verified all AI output.

\begin{figure*}
  \centering
  \includegraphics[width=\textwidth]{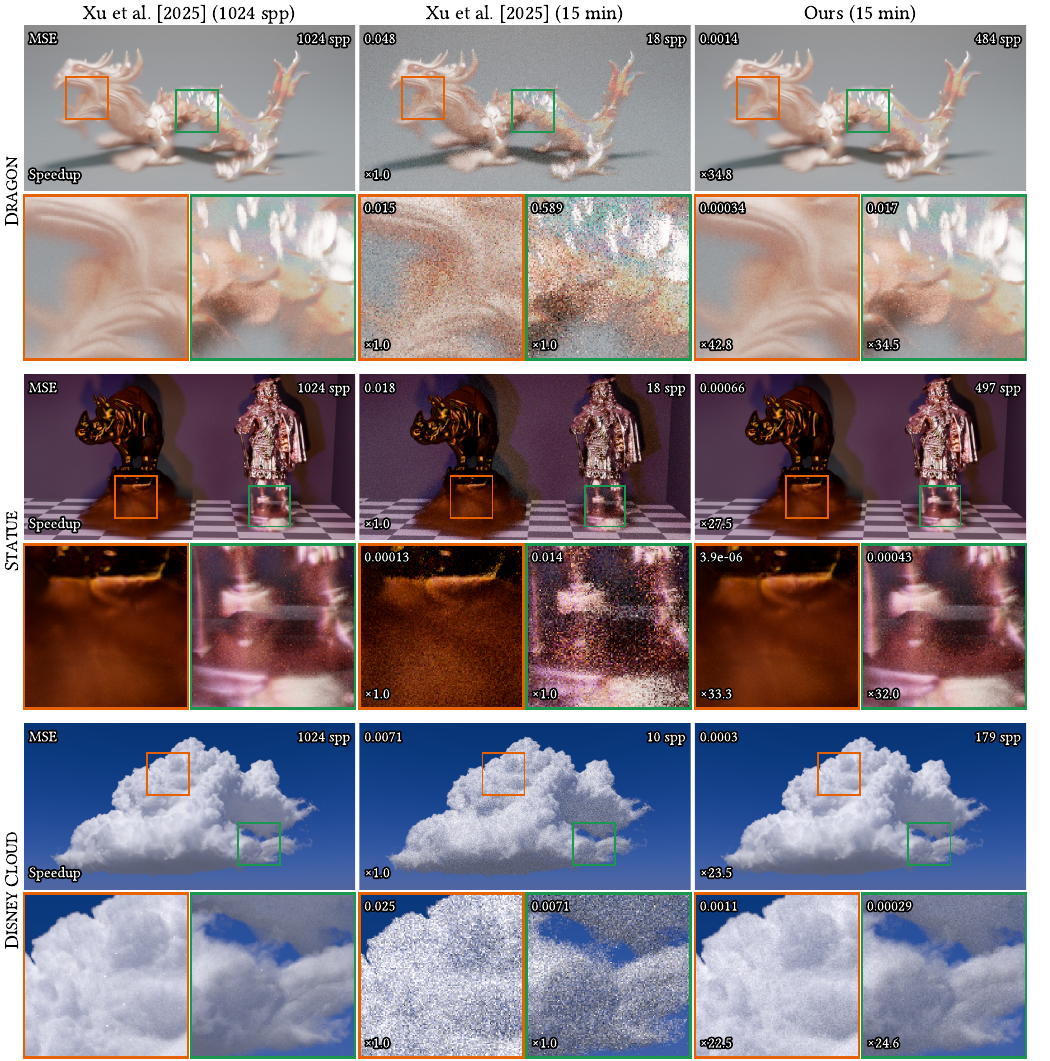}
  \caption{Equal-time GPIS rendering on \textsc{Dragon},
    \textsc{Statue}, and \textsc{Disney Cloud}.
    The left column shows 1024-spp renderings from the 1D
    sparse-convolution method of \citet{Xu:2025:Practical}
    with renewal-half+ conditioning.
    The middle and right columns compare that method and ours,
    respectively, after 15\,min per scene.
    Each method's MSE is measured against its own converged
    rendering.
    Annotations report sample counts and MSEs, with
    ``speedup'' denoting the MSE ratio relative to the
    sparse-convolution baseline for each image or marked region.}
  \label{fig:equal_time}
\end{figure*}

\subsection{Forward Rendering}
\Cref{fig:equal_time} compares our method with
sparse-convolution GPIS using renewal-half+
conditioning \citep{Xu:2025:Practical}, with equal
15\,min budgets per scene.
This baseline substantially outperforms the function-space
sampling of \citet{Seyb:2024:Microfacets}, which we
therefore omit.

\textsc{Dragon} and \textsc{Statue} use nonstationary
squared-exponential kernels.
For \textsc{Disney Cloud}, we approximate the original
Henyey--Greenstein phase function ($g=0.877$) by isotropic
scattering using the first-order similarity relation
\citep{Zhao:2014:Highorder}, scaling the target extinction
by $(1-g)$.
We then convert this medium to GPIS parameters using the
heterogeneous inversion of \cref{sec:inversion_heterog}.

The two methods use different conditioning approximations
and therefore converge to slightly different images.
We measure each method's MSE against its own converged
render, so the reported ratios measure estimator
efficiency.
Full-resolution renders are provided in the
supplemental for comparing the two transport
models.

By avoiding realization synthesis and per-path conditioning,
our method achieves $23.5$--$34.8\times$ lower MSE across
the three scenes at equal time.
The \textsc{Statue} scene also shows that the same medium
can transition continuously between hard-surface and
volumetric appearance within a single object.

\subsection{Surface Scattering}

\begin{figure*}
  \centering
  \includegraphics[width=\textwidth]{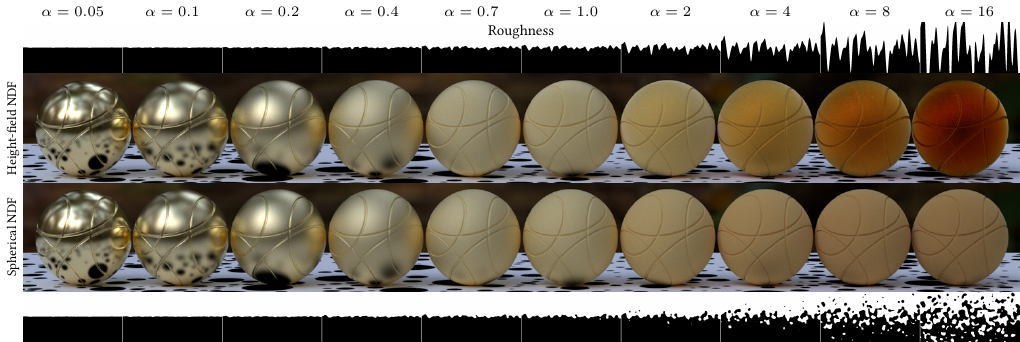}
  \caption{As roughness $\alpha$ of a gold Beckmann BRDF increases (left to right) the corresponding microsurface of the heightfield becomes unreasonably spiky and the BRDF strongly saturates and darkens (top right).  Our new spherical Beckmann NDF offers a close match for low roughness (left) but offers a stable high roughness limit (bottom right) by modeling porous materials, at negligible extra rendering cost.}
  \label{fig:high_roughness}
\end{figure*}

\Cref{fig:high_roughness} compares the geometry and multiply scattered
appearance of height-field and full-sphere Beckmann models across
increasing roughness. For this comparison, we instantiate the
full-sphere family with $\CondGradMean=\hat{\mathbf{z}}$ and
$\CondGradCov=(\alpha^2/2)\mathbf{I}$, providing an isotropic
one-parameter slice analogous to classical Beckmann. The models behave
similarly at low roughness but diverge as roughness increases: the
height field develops increasingly sharp spikes, whereas the
full-sphere model transitions toward porous geometry. At high
roughness, the full-sphere model avoids the extreme darkening exhibited
by its height-field counterpart.

\begin{figure}
  \centering
  \includegraphics[width=\linewidth]{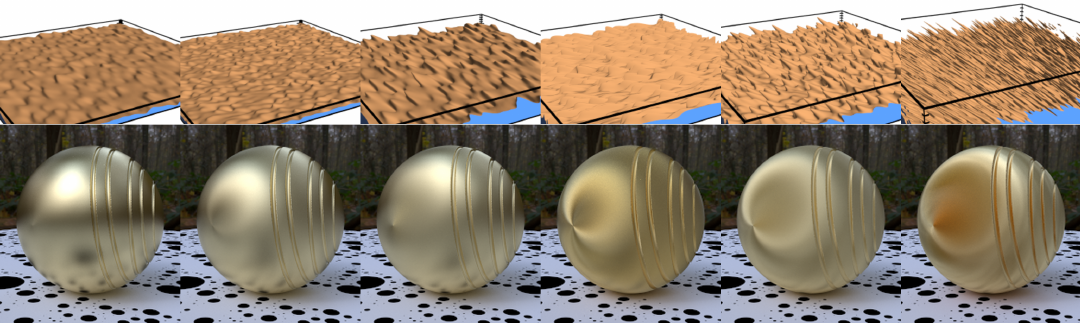}
  \caption{Full-sphere GPIS BRDFs under varying $\Sigma_{\bar{g}}$ with visualizations of microsurface geometry (top) and multiply scattered appearance (bottom). The columns span isotropic scattering, in-plane anisotropy, and correlations between the lateral and depth gradient components, which tilt the anisotropy out of the mean surface plane.}
  \label{fig:brdf_tilted}
\end{figure}

Whereas \cref{fig:high_roughness} uses an isotropic covariance,
\cref{fig:brdf_tilted} demonstrates the broader family enabled by a
full $\CondGradCov$, comparing isotropic scattering with in-plane and
out-of-plane anisotropy.

\begin{figure}
  \centering
  \includegraphics[width=\linewidth]{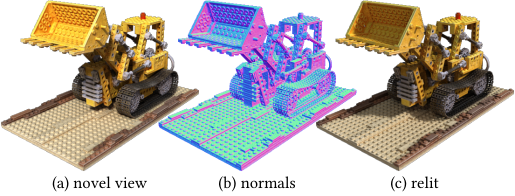}
  \caption{A lifted radiance field is a renderable asset.
        We train TensoRF~\citep{Chen:2022:TensoRF} on all 100
        training views of the \textsc{Lego} scene and convert its
        density grid to GPIS parameters.
        (a)~The mean surface, with the field's view-averaged color
        used as diffuse albedo.
        (b)~The normal map of the mean surface.
        (c)~The same asset under new illumination. Relighting is
        approximate because the baked color retains shading from
        the training illumination.
        No mesh extraction, second reconstruction stage, or
        additional inverse rendering is required.}
  \label{fig:lego_asset}
\end{figure}

\subsection{Inverting Radiance Fields}

\paragraph{From density grid to geometry and normals}
\Cref{fig:lego_asset} shows the geometry and shading normals
recovered from a trained radiance field.
We train TensoRF~\citep{Chen:2022:TensoRF} on all 100 views
of the \textsc{Lego} scene and convert its density grid to
GPIS parameters using the pointwise inversion of
\cref{sec:inversion_heterog}.
The void level defines the mean surface, and the direction
of $\CondGradMean$ gives its shading normals.
This yields a renderable asset without mesh extraction or
further reconstruction.

We also bake the field's view-averaged color as diffuse
albedo and render the asset under modified illumination.
This relighting is approximate because the baked color
retains shading from the training illumination.
Recovering illumination-independent albedo would require
an additional decomposition
\citep{Zhang:2021:NeRFactor,Boss:2021:NeRD,Jin:2023:TensoIR}.

\begin{figure*}
  \centering
  \includegraphics[width=\textwidth]{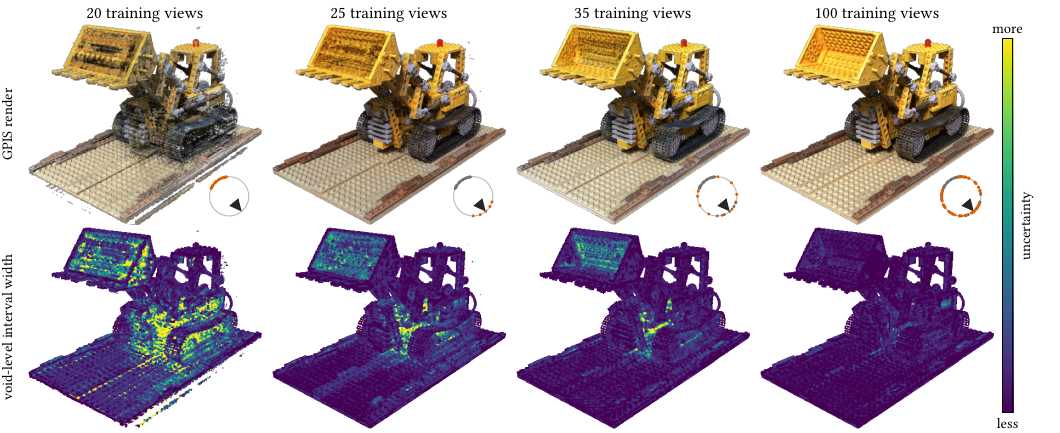}
  \caption{Converting radiance fields to GPISes visualizes
reconstruction ambiguity.
We train TensoRF~\citep{Chen:2022:TensoRF} on growing
subsets of the 100 \textsc{Lego} views and convert each
density grid to GPIS parameters.
Left to right: 20 views opposite the camera, five added
facing the camera, ten added around the ring, and all 100.
Ring insets show training-view azimuths; orange marks
newly added views, and the camera icon marks the fixed
render viewpoint.
Top row: GPIS renderings, with $\VoidLevel>+2$ treated as
vacuum to remove floaters.
Bottom row: total ray length over which $\VoidLevel$
lies between $-2$ and $+2$, shown on a shared color scale.
White indicates no density along the ray.
Additional views reduce the measured length in the regions
they constrain. This visualizes ambiguity in the
reconstructed density, without providing calibrated
uncertainty.}
  \label{fig:rf_uncertainty}
\end{figure*}

\paragraph{Rendering reconstruction uncertainty}
In \cref{fig:rf_uncertainty}, we apply the same conversion
to TensoRF reconstructions trained on increasing subsets
of the 100 \textsc{Lego} views.
Regions with limited view coverage have a volumetric
appearance after conversion.
To visualize this ambiguity, we measure the total length
along each camera ray over which
$\VoidLevel[\mathbf{x}]$ lies between $-2$ and $+2$.
Starting from 20 views opposite the camera, adding five
views facing the camera reduces this length at the front;
ten more views around the ring reduce it along the flanks;
and all 100 views reduce it nearly to zero.

The conversion exposes ambiguity encoded in the
reconstructed density, without providing calibrated
uncertainty.
It inherits the model's errors: a confidently incorrect
reconstruction can still yield a small measured length.

\section{Discussion, Limitations, and Conclusion}
\label{sec:discussion}

\paragraph{Comparison to \citet{Huang:2026:Macrofacet}}
Our derivation begins with general GP statistics, whereas
\citet{Huang:2026:Macrofacet} extend microfacet theory into
a volume around a macro-surface.
Their assumption that local field and gradient values are
independent of those at the ray origin plays a similar role
to our local-conditioning approximation.
For their linear-ramp mean and stationary, axis-aligned SE
kernel, the extinction coefficient, NDF, and vNDF coincide
algebraically with ours; the phase functions also agree
for the same conductor micro-BRDF.

Our formulation handles nonstationary means and covariances,
including the conditional-gradient corrections in
\cref{eqn:condgrad_identities} and tilted anisotropy.
We analytically establish the projected-area normalization
\eqref{eqn:projarea_def} and NDF first-moment identity
\eqref{eqn:ndf_first_moment}, and recover Beckmann in the
height-field limit (\cref{app:beckmannlimit}).
Our gradient-space construction samples the vNDF exactly
(\cref{alg:samplevndf}), while their sampler uses a mixture
of Beckmann-vNDF and uniform-hemisphere proposals.
We further derive masking and shadowing for full-sphere
Beckmann and GGX surface models
(\cref{sec:surfaceconnections}) and construct
the inverse map from RTE parameters to GPIS specifications
(\cref{sec:inversion}).

\paragraph{Memory and correlation}
The main approximation in our construction is local conditioning:
at each prospective collision, we condition on the current point
being in the void and discard earlier ray observations.
In homogeneous media, this yields exponential free flights,
whereas those of the underlying GPIS are generally non-exponential.
Our comparisons show larger discrepancies in concavities narrower
than the correlation length.
Agreement generally improves when the mean void chord is long
relative to that length, so crossings are sparse on the scale
of the correlation.
Non-exponential transport
\citep{Larsen:2011:Generalized,Bitterli:2018:Radiative,Jarabo:2018:Radiative}
and Gaussian first-passage models
\citep{Roberts:1999:Chorddistribution} offer possible ways to restore
part of this missing history.

\paragraph{Scope of the inverse map}
Our inverse map exactly matches pointwise LCA coefficients for
homogeneous isotropic and SGGX media and compatible heterogeneous
targets (\cref{sec:inversion_homog,sec:inversion_heterog}).
Extinction and scattering must be jointly representable by our
GPIS NDF families and chosen micro-BRDF, ensuring consistency
between attenuation and the underlying surface-normal statistics.
The heterogeneous conversion used in our results is approximate:
spatial variation in $\VoidLevel$ induces NDF asymmetry and
directional extinction \eqref{eqn:isotropy_error}.
The supplemental construction approximately matches spatially
varying covariance statistics when parameters vary slowly over
a correlation length.
Future work could develop nonstationary covariance kernels that
exactly reproduce compatible spatially varying extinction and
scattering under the LCA, without requiring parameters to vary
slowly over a correlation length.

\paragraph{Two-sided micro-BSDFs and Extensions}
The natural micro-BSDF for a GPIS separating void from solid matter is an opaque
surface model (mirror or Lambertian), but in principle any physically valid $f_r$ may
replace the mirror in \cref{eqn:phasefn}.
In the dilute regime (\cref{sec:inversion})---where $\GPField < 0$ regions are rare
and small, such as water droplets suspended in air---a dielectric micro-BSDF or
Lorenz--Mie micro-phase function could summarize the scattering (and negligible traversal) within each
solid particle.
If macroscopic transport within the $\GPField < 0$ regions is desired, a stateful
extension is needed.
Up-crossings---events at which the field increases through zero ($\DirDeriv[t] > 0$)
---are governed by $\ProjAreaPlus[\RayDir]$ \eqref{eqn:projarea_plus}.
By the same local-conditioning argument as \cref{sec:kacrice}, conditioned on
$\GPField[t] < 0$, the up-crossing extinction coefficient is
\begin{align}
  \label{eqn:chi_plus}
  \ExtCoeffP[\mathbf{r}(t), \RayDir]
    = \frac{\normPDF[\MeanF[t]/\StdF[t]]}
           {\StdF[t]\;\normCDF[-\MeanF[t]/\StdF[t]]}
      \cdot \ProjAreaPlus[\mathbf{r}(t), \RayDir],
\end{align}
parallel to \cref{eqn:extinction_coeff} but with
$\mathbb{P}(\GPField[t] < 0)$ in the denominator.
Transmission toggles the medium phase; reflection preserves it.
Subsequent free flights use $\ExtCoeff$ in the void and
$\ExtCoeffP$ in the solid.

\paragraph{Differentiable rendering and reconstruction}
Our conversion provides a route to optimizing GPIS parameters
using existing volumetric differentiable-rendering methods.
The void level, gradient statistics, and micro-BSDF parameters
encode a probabilistic surface, its local orientation and
normal variation, and its material response.
Jointly reconstructing these properties is one promising
direction for future work.

\paragraph{Conclusion}
We connect GPISes and participating media through an anisotropic
transport model with closed-form extinction and NDFs under the
local-conditioning approximation.
Our full-sphere Beckmann and GGX families support exact
visible-normal sampling, analytic masking--shadowing, and
single-scattering models for specular microsurfaces,
with extensions to multiple scattering.
The framework recovers SGGX in the centered case and
Beckmann--Smith in the height-field limit.
Our realization-free renderer achieves $23$--$34\times$
lower equal-time estimator MSE than the realization-based
baseline.
The inverse map constructs GPISes from compatible volumetric
assets and radiance-field reconstructions.

\printbibliography

\appendix
\crefalias{section}{appendix}%
\crefalias{subsection}{subappendix}%

\section{Derivation of the NDF Radial Integral}
\label{app:ndfintegral}

Using the scalars $\NDFA,\NDFB,\NDFC$ from
\cref{eqn:ndf_abc}, with dependence on $\SurfNorm$ suppressed,
the radial integral in \cref{eqn:ndf_def}, before Gaussian
normalization, is
\begin{align}
  I(\SurfNorm)
    = \int_0^\infty r^3
      \exp\!\left[-\tfrac12
        \left(\NDFA r^2-2\NDFB r+\NDFC\right)\right]\dif r.
\end{align}
Completing the square and substituting
$u=\sqrt{\NDFA}\,r-\NDFb[]$, where
$\NDFb[]=\NDFB/\sqrt{\NDFA}$, gives
\begin{align}
  \label{eqn:Itilde}
  I(\SurfNorm)
    = \frac{\euler^{(\NDFb[][2]-\NDFC)/2}}{\NDFA[][2]}
      \int_{-\NDFb[]}^\infty
        (u+\NDFb[])^3\euler^{-u^2/2}\,\dif u.
\end{align}

Expanding the cubic requires the following truncated Gaussian
integrals. The last two follow by integration by parts using
$\dif(\euler^{-u^2/2})=-u\euler^{-u^2/2}\,\dif u$:
\begin{align*}
  \int_{-\NDFb[]}^\infty \euler^{-u^2/2}\,\dif u
    &= \sqrt{2\pi}\,\normCDF[\NDFb[]], \\
  \int_{-\NDFb[]}^\infty u\euler^{-u^2/2}\,\dif u
    &= \sqrt{2\pi}\,\normPDF[\NDFb[]], \\
  \int_{-\NDFb[]}^\infty u^2\euler^{-u^2/2}\,\dif u
    &= \sqrt{2\pi}\Bigl[
         \normCDF[\NDFb[]]
         -\NDFb[]\normPDF[\NDFb[]]
       \Bigr], \\
  \int_{-\NDFb[]}^\infty u^3\euler^{-u^2/2}\,\dif u
    &= \sqrt{2\pi}(\NDFb[][2]+2)\normPDF[\NDFb[]].
\end{align*}
Substituting these moments and collecting terms gives
\begin{align}
  \label{eqn:step4}
  &\int_{-\NDFb[]}^\infty
    (u+\NDFb[])^3\euler^{-u^2/2}\,\dif u
    \notag\\
  &\qquad =
    \sqrt{2\pi}\Bigl[
      (\NDFb[][2]+2)\normPDF[\NDFb[]]
      +\NDFb[](\NDFb[][2]+3)\normCDF[\NDFb[]]
    \Bigr].
\end{align}
Substituting into \cref{eqn:Itilde} and restoring the Gaussian
normalization $(2\pi)^{-3/2}|\CondGradCov|^{-1/2}$
from \cref{eqn:gaussian_pdf} gives \cref{eq:ndf}.

\section{Masking and Shadowing for BRDF-Compatible GPISes}
\label{app:heightfield-masking}

We derive the marginal masking function $\ShadowG[\OutDir]$
and the bistatic masking-shadowing function
$G_{12}(\OutDir[i],\OutDir[o])$ from directional transmittances
for both the Gaussian and GGX families.

\paragraph{One-sided projected areas}
For a BRDF-compatible GPIS, $\CondGradMean=\hat{\mathbf{z}}$,
and the precision mixture of \cref{sec:mixture} preserves this mean.
Thus, along an outward direction $\OutDir$,
$\CondDerivMean=\OutDir\cdot\hat{\mathbf{z}}=\cos\theta_\ast>0$.

Since $\dif z/\dif t=\cos\theta_\ast$, changing variables
from ray length to height in the extinction integral divides
the projected-area factor by $\cos\theta_\ast$.
We therefore define Smith's auxiliary function as
\begin{equation}
  \SmithLambda[\OutDir]
    \coloneq \frac{\ProjAreaMinus[\OutDir]}{\cos\theta_\ast}.
  \label{eqn:brdf_lambda_definition}
\end{equation}
As shown below, this is the exponent in the outgoing
transmittance.

Set $\nu=\cos\theta_\ast/\CondDerivStd$.
Substituting the Gaussian and heavy-tailed projected areas
from \cref{eqn:one_sided_area_closed,eqn:mixture_projarea}
into \cref{eqn:brdf_lambda_definition} gives
\begin{align}
  \SmithLambda[\OutDir]
    =
    \begin{cases}
      \displaystyle
      \frac{\normPDF[\nu]}{\nu}-\normCDF[-\nu],
        & \text{Gaussian},\\[6pt]
      \displaystyle
      \frac{\sqrt{1+2/\nu^2}-1}{2},
        & \text{GGX}.
    \end{cases}
  \label{eqn:brdf_lambda_closed_forms}
\end{align}
For GGX, $\CondDerivStd$ denotes the scale before mixing.
In either case,
\begin{align}
  \ProjAreaMinus[\OutDir]
    &= \cos\theta_\ast\,\SmithLambda[\OutDir].
  \label{eqn:brdf_outward_area}
\end{align}
Since mixing preserves the mean gradient, the asymmetry
identity \eqref{eqn:asymmetry} holds for both families.
For the opposite direction, it gives
\begin{align}
  \ProjAreaMinus[-\OutDir]
    &= \ProjAreaMinus[\OutDir]+\cos\theta_\ast \notag\\
    &= \cos\theta_\ast
       \bigl[1+\SmithLambda[\OutDir]\bigr].
  \label{eqn:brdf_inward_area}
\end{align}

\paragraph{Directional transmittances}
Set $u=\normCDF[z/\StdF]$.
As specified in \cref{sec:surfacelimit}, both families use
the height-dependent occupancy factor of
\cref{eqn:extinction_coeff} with their respective,
spatially constant projected areas:
\begin{align}
  \ExtCoeff[z,\RayDir]
    = \frac{\normPDF[z/\StdF]}{\StdF\,u}\,
      \ProjAreaMinus[\RayDir].
  \label{eqn:brdf_extinction_factorization}
\end{align}

For an outward ray $\mathbf{r}(t)$ starting at height $z$,
transmittance follows by integrating extinction over ray length:
\begin{align}
  \Trans[z\to\infty,\OutDir]
    = \exp\!\left[
        -\int_0^\infty
          \ExtCoeff[\mathbf{r}(t),\OutDir]\,\dif t
      \right].
  \label{eqn:brdf_transmittance_integral}
\end{align}
Changing variables using
$\dif z'=\cos\theta_\ast\,\dif t$ and
$u'=\normCDF[z'/\StdF]$, with
$\dif u'=\normPDF[z'/\StdF]\,\dif z'/\StdF$,
the occupancy integral becomes
\begin{align}
  \int_z^\infty
    \frac{\normPDF[z'/\StdF]}
         {\StdF\,\normCDF[z'/\StdF]}\,\dif z'
    = \int_u^1 \frac{\dif u'}{u'}
    = -\ln u.
  \label{eqn:brdf_occupancy_integral}
\end{align}
Substituting this result and using
\cref{eqn:brdf_outward_area} gives
\begin{align}
  \Trans[z\to\infty,\OutDir]
    &= \exp\!\left[
         \frac{\ProjAreaMinus[\OutDir]}{\cos\theta_\ast}
         \ln u
       \right] \notag\\
    &= u^{\SmithLambda[\OutDir]}.
  \label{eqn:brdf_outgoing_transmittance}
\end{align}

For the incoming direction $-\OutDir$,
$\dif z'=-\cos\theta_\ast\,\dif t$ reverses the integration
bounds. The same occupancy integral therefore gives
\begin{align}
  \Trans[\infty\to z,-\OutDir]
    = u^{\ProjAreaMinus[-\OutDir]/\cos\theta_\ast}
    = u^{1+\SmithLambda[\OutDir]},
  \label{eqn:brdf_incoming_transmittance}
\end{align}
using \cref{eqn:brdf_inward_area}.

\paragraph{Marginal masking}
The outgoing transmittance gives an escape probability that
depends on both height and direction. To obtain a masking
function that depends only on direction, we average this
probability over all crossing heights. For a fixed direction,
the Kac--Rice rate
$p_{\GPField}(0)\ProjAreaMinus[-\OutDir]$
varies with height only through $p_{\GPField}(0)$,
so this density supplies the relative weights in the average.
Using
\begin{align}
  p_{\GPField}(0)\,\dif z
    = \frac{\normPDF[z/\StdF]}{\StdF}\,\dif z
    = \dif u,
  \label{eqn:brdf_crossing_height_measure}
\end{align}
we obtain
\begin{align}
  \ShadowG[\OutDir]
    &= \int_{-\infty}^{\infty}
         \Trans[z\to\infty,\OutDir]\,
         p_{\GPField}(0)\,\dif z \notag\\
    &= \int_0^1 u^{\SmithLambda[\OutDir]}\,\dif u
     = \frac{1}{1+\SmithLambda[\OutDir]}.
  \label{eqn:brdf_marginal_masking}
\end{align}

\paragraph{Bistatic masking}
Under the local-conditioning approximation, we model bistatic
distant masking by treating the two escape paths as conditionally
independent given their shared crossing height. This neglects
directional visibility correlations between the paths. We average
the product of their transmittances over crossing heights.
\begin{align}
  G_{12}(\OutDir[i],\OutDir[o])
    &= \int_0^1
         u^{\SmithLambda[\OutDir[i]]}\,
         u^{\SmithLambda[\OutDir[o]]}\,\dif u \notag\\
    &= \frac{1}
         {1+\SmithLambda[\OutDir[i]]
            +\SmithLambda[\OutDir[o]]}.
  \label{eqn:brdf_bistatic_masking}
\end{align}
The shared height couples the two directions through this
average, recovering \cref{eqn:brdf_single_height_integral}.

\paragraph{Checks}
The symmetry of $G_{12}(\OutDir[i],\OutDir[o])$ under
exchange of its arguments is consistent with BRDF reciprocity.
As $\SmithLambda[\OutDir[o]]\to0$,
\cref{eqn:brdf_bistatic_masking} reduces to
$\ShadowG[\OutDir[i]]$, so an unobstructed outgoing
direction introduces no additional shadowing.
\section{GPIS height-field limit}
\label{app:ndfspecialcases}

We examine the height-field limit of the GPIS NDF,
one-sided projected area, and transmittance. The NDF reduces to
the Beckmann distribution, while the projected area and
transmittance under local conditioning recover Smith's marginal
and height-conditional masking functions, respectively.

\paragraph{Height-field setup}
Let $\GPField[x,y,z] = z - \HeightField[x,y]$, where $\HeightField$ is a stationary,
zero-mean 2D GP with height variance $\HeightStd^2$, spatial covariance
$\CovFnHF[\Delta x][\Delta y]$, and slope variance $\SlopeStd^2 = -\CovFnHF''(0)$.
The field mean is $\GPMean = z$, the field variance $\VarF = \HeightStd^2$ is constant,
and stationarity forces the cross-covariance $\CovFFp = 0$.
The conditional gradient statistics therefore equal their unconditional counterparts:
$\CondGradMean = \nabla\GPMean = \hat{\mathbf{z}}$ and $\CondGradCov = \CovGrad$.
We parameterize the gradient covariance as
\begin{align}
  \label{eqn:hf_gradcov}
  \CondGradCov = \mathrm{diag}(\epsilon_h,\, \epsilon_h,\, \epsilon_z),
\end{align}
where $\epsilon_h = \SlopeStd^2$ is the horizontal gradient variance and
$\epsilon_z \to 0$ in the height-field limit.

\subsection{Beckmann NDF}
\label{app:beckmannlimit}

We show that \cref{eq:ndf} reduces to the Beckmann distribution: for
$\cos\theta > 0$ it converges to the Beckmann density of
\cref{eqn:beckmann_limit}, and for $\cos\theta \le 0$ it vanishes,
recovering the one-sided support. We look at the upper hemisphere first.

\paragraph{Asymptotic evaluation of the NDF scalars}
In
standard spherical coordinates $\SurfNorm = (\sin\theta\cos\phi,\, \sin\theta\sin\phi,\, \cos\theta)$, and the quadratic-form scalars from \cref{eqn:ndf_abc} become:
\begin{align}
  \NDFA &= \frac{\sin^2\theta}{\epsilon_h} + \frac{\cos^2\theta}{\epsilon_z}, \quad
  \NDFB  = \frac{\cos\theta}{\epsilon_z}, \quad
  \NDFC      = \frac{1}{\epsilon_z}.
\end{align}

For $\cos\theta > 0$ (upper hemisphere), substituting the scalars above gives
\begin{align}
  \NDFb[][2]-\NDFC
    &= \frac{\NDFB[][2]}{\NDFA}-\NDFC \notag\\
    &= \frac{1}{\epsilon_z}
       \left[
         \frac{\epsilon_h\cos^2\theta}
              {\epsilon_h\cos^2\theta+\epsilon_z\sin^2\theta}
         -1
       \right] \notag\\
    &= -\frac{\sin^2\theta}
             {\epsilon_h\cos^2\theta+\epsilon_z\sin^2\theta}
       \xrightarrow{\epsilon_z\to0}
       -\frac{\tan^2\theta}{\epsilon_h}.
  \label{eqn:bsq_minus_C}
\end{align}
Combining the terms in brackets gives a numerator of
$-\epsilon_z\sin^2\theta$, whose $\epsilon_z$ factor cancels
the prefactor $1/\epsilon_z$.
The exponential factor in \cref{eq:ndf} therefore tends to
$\euler^{-\tan^2\theta/(2\epsilon_h)}$.

\paragraph{Dominant term in the bracket}
As $\epsilon_z \to 0$, $b = B/\sqrt{A} \approx 1/\sqrt{\epsilon_z} \to +\infty$ for
$\cos\theta > 0$.
In the bracket of \cref{eq:ndf}, the term $\NDFb[](\NDFb[][2]+3)\normCDF[\NDFb[]] \to b^3$
dominates (since $\normCDF[b] \to 1$), while
$(\NDFb[][2]+2)\normPDF[\NDFb[]]$ decays exponentially.

\paragraph{$\epsilon_z$ cancellation}
The leading factors in \cref{eq:ndf} are
$|\CondGradCov|^{1/2}=\epsilon_h\sqrt{\epsilon_z}$,
$\NDFA[][2]\approx\cos^4\theta/\epsilon_z^2$, and
$\NDFb[][3]\approx\epsilon_z^{-3/2}$.
Their powers of $\epsilon_z$ cancel on substitution, giving
\begin{align}
  \NDF[\SurfNorm]
    \xrightarrow{\epsilon_z\to0}
    \frac{\euler^{-\tan^2\theta/(2\epsilon_h)}}
         {2\pi\,\epsilon_h\,\cos^4\theta}.
  \label{eqn:beckmann_result}
\end{align}
This is the Beckmann NDF \eqref{eqn:beckmann_limit}
with roughness $\alpha=\sqrt{2\epsilon_h}$.

\paragraph{Lower hemisphere}
We show that the NDF vanishes for $\cos\theta\le0$ by
bounding it above by a quantity that tends to zero.
For these directions, $\NDFb[]\le0$, so the CDF contribution
$\NDFb[](\NDFb[][2]+3)\normCDF[\NDFb[]]$ is nonpositive.
Dropping this term from \cref{eq:ndf} and combining the
remaining Gaussian density with the exponential factor gives
\begin{align}
  0\le\NDF[\SurfNorm]
    &\le
    \frac{(\NDFb[][2]+2)\,\euler^{-1/(2\epsilon_z)}}
         {(2\pi)^{3/2}\,|\CondGradCov|^{1/2}\,\NDFA[][2]}.
  \label{eqn:hf_lower_hemisphere_bound}
\end{align}
The lower bound follows from the nonnegative integral
defining $\NDF$ in \cref{eqn:ndf_def}.
The prefactor on the right grows at most as a power of
$1/\epsilon_z$, while the exponential decays faster than
any such power. Thus the upper bound tends to zero,
forcing $\NDF[\SurfNorm]\to0$ for $\cos\theta\le0$
and recovering the one-sided Beckmann support. $\square$

\subsection{One-sided projected area and Smith masking}
\label{app:beckmannsmith}
We now show that our local-conditioning approximation reduces
to Smith's local shadowing approximation in the height-field
limit, and that the one-sided projected area recovers Smith's
masking function for Gaussian slopes.

For a height field, $\GPField>0$ is equivalent to
$\HeightField<z$. Thus \cref{eqn:local_conditioning} replaces
conditioning on the full ray history by conditioning on the
surface lying below the current ray point, with normalization
$\mathbb{P}(\HeightField<z)=\normCDF[z/\HeightStd]$.
This is Smith's local shadowing assumption
\citep{Smith:1967:Geometrical}.
We obtain the corresponding projected-area result by
specializing \cref{app:heightfield-masking}.

For an outward direction $\OutDir$, substituting
\cref{eqn:hf_gradcov} into \cref{eqn:ray_cond_stats} gives
$\CondDerivMean=\cos\theta_\ast$ and
$\CondDerivVar
=\epsilon_h\sin^2\theta_\ast+\epsilon_z\cos^2\theta_\ast
\to\SlopeStd^2\sin^2\theta_\ast$.
The standardized derivative therefore becomes
$\SlopeParam=\cot\theta_\ast/\SlopeStd$ in the height-field
limit. Using \cref{eqn:brdf_inward_area,eqn:brdf_marginal_masking},
we obtain
\begin{align}
  \ProjAreaMinus[-\OutDir]
    \xrightarrow{\epsilon_z\to0}
    \cos\theta_\ast\bigl[1+\SmithLambda[\SlopeParam]\bigr]
    = \frac{\cos\theta_\ast}
           {\ShadowG^{\mathrm{Smith}}(\theta_\ast)},
  \label{eqn:hf_projarea}
\end{align}
where
$\ShadowG^{\mathrm{Smith}}(\theta_\ast)
=1/[1+\SmithLambda[\SlopeParam]]$
is the Smith masking function for Gaussian slopes.

\subsection{Transmittance and Smith height-conditional masking}
\label{app:smith-height-conditional-masking}

We now show that the GPIS transmittance under local conditioning
exactly recovers Smith's height-conditional masking function
in the height-field limit.

Setting $\StdF=\HeightStd$ in
\cref{eqn:brdf_outgoing_transmittance}, the empty-space
probability coordinate becomes
$u=\normCDF[z/\HeightStd]$, the surface-height CDF.
The outgoing escape probability therefore gives
\begin{align}
  \ShadowG[z][\OutDir]
    = \Trans[z\to\infty,\OutDir]
    = \bigl[\normCDF[z/\HeightStd]\bigr]^{
        \SmithLambda[\SlopeParam]},
  \label{eqn:hf_height_conditional_masking}
\end{align}
which is Smith's height-conditional masking function
\citep{Smith:1967:Geometrical,Heitz:2016:Multiplescattering}.
Averaging over height as in \cref{eqn:brdf_marginal_masking}
recovers the marginal Smith shadowing function
$\ShadowG^{\mathrm{Smith}}(\theta_\ast)
=1/[1+\SmithLambda[\SlopeParam]]$.
The same height-field substitution in
\cref{eqn:brdf_bistatic_masking} recovers Smith's bistatic
masking-shadowing function
\citep[Eq.~21]{Dupuy:2016:Additional}. $\square$

\section{Heavy-Tailed NDF, Projected Area, and Sampling}
\label{app:mixture}
This appendix derives \cref{eqn:mixture_ndf,eqn:mixture_projarea} of
\cref{sec:mixture} for general $(\CondGradMean,\CondGradCov)$.

\paragraph{Gradient distribution}
Applying the precision mixture $\tau \sim \mathrm{Exp}(1)$,
$\GradField\mid\tau \sim \mathcal{N}(\CondGradMean,\,\CondGradCov/\tau)$, and
marginalizing over $\tau$ using
$\int_0^\infty \tau^{3/2}\euler^{-\tau(1+Q/2)}\dif\tau = \Gamma(5/2)(1+Q/2)^{-5/2}$, with
$Q(\GradField) \coloneq (\GradField-\CondGradMean)^\mathsf{T}\CondGradCov^{-1}(\GradField-\CondGradMean)$
and $\Gamma(5/2)=3\sqrt{\pi}/4$, gives
\begin{align}
  \label{eqn:ggx_grad}
  p^{\mathrm{GGX}}_{\CondGrad}(\GradField)
    = \frac{3\sqrt{\pi}}{4(2\pi)^{3/2}|\CondGradCov|^{1/2}}
      \Bigl(1+\tfrac12 Q(\GradField)\Bigr)^{-5/2}.
\end{align}

\paragraph{NDF integral}
Substituting $\GradField = r\SurfNorm$ into \cref{eqn:ggx_grad} and using
$Q(r\SurfNorm) = \NDFA[\SurfNorm] r^2 - 2\NDFB[\SurfNorm] r + \NDFC$ from
\cref{eqn:ndf_abc}, complete the square with
$\mu \coloneq \NDFB[\SurfNorm]/\NDFA[\SurfNorm] = \NDFb[]/\sqrt{\NDFA[]}$:
\begin{align}
  1+\tfrac12 Q(r\SurfNorm) = \frac{\NDFA[]}{2}(r-\mu)^2 + k, \quad
  k \coloneq 1+\frac{\NDFC-\NDFb[][2]}{2}.
\end{align}
Writing $\Delta \coloneq 2k/\NDFA[]$ and rescaling $r = \mu+\sqrt{\Delta}\,t$
reduces the radial integral to the dimensionless form
\begin{align}
  \int_{-\lambda}^\infty(\lambda+t)^3(1+t^2)^{-5/2}\dif t, \quad
  \lambda \coloneq \mu/\sqrt{\Delta},
\end{align}
which evaluates to $(2R-\lambda)/(3(R-\lambda)^2)$ with $R=\sqrt{1+\lambda^2}$ (verified
by differentiation, or by checking $\lambda=0$ against the elementary substitution
$w=1+t^2$, which gives $2/3$, matching $(2R-\lambda)/(3(R-\lambda)^2)$ at $\lambda=0$).
Undoing the rescaling gives
\begin{align}
  \int_0^\infty r^3\bigl((r-\mu)^2+\Delta\bigr)^{-5/2}\dif r
    &= \frac{2S-\mu}{3(S-\mu)^2}, \\
  S &\coloneq \sqrt{\mu^2+\Delta} = \sqrt{(\NDFC+2)/\NDFA[]}.
\end{align}
Substituting into \cref{eqn:ggx_grad}, whose prefactor contributes an additional
factor of $(\NDFA[]/2)^{-5/2}$, gives us \cref{eqn:mixture_ndf} from the main text.
\hfill$\square$

\paragraph{One-sided projected area}
Linear projections of the mixture remain in the same family one dimension down:
$\CondDeriv\mid\tau \sim \mathcal{N}(\CondDerivMean,\CondDerivVar/\tau)$, so the same
$\Gamma(3/2,\cdot)$ integral used above, now with $\Gamma(3/2)=\sqrt{\pi}/2$, gives
\begin{align}
  p_{\CondDeriv}^{\mathrm{GGX}}(v)
    = \frac{\alpha_{\mathrm{eff}}^2/2}{\bigl((v-\CondDerivMean)^2+\alpha_{\mathrm{eff}}^2\bigr)^{3/2}},
  \quad \alpha_{\mathrm{eff}}^2 \coloneq 2\,\CondDerivVar.
\end{align}
Evaluating
$\ProjAreaMinus[\RayDir]^{\mathrm{GGX}} = \int_{-\infty}^0 |v|\,p_{\CondDeriv}^{\mathrm{GGX}}(v)\dif v$
via the substitution $s=v-\CondDerivMean$ and the antiderivative
$\int s\,(s^2+\alpha_{\mathrm{eff}}^2)^{-3/2}\dif s = -(s^2+\alpha_{\mathrm{eff}}^2)^{-1/2}$
gives us \cref{eqn:mixture_projarea} from the main text. \hfill$\square$

\subsection{Exact vNDF Sampling}
\label{app:mixture_sampling}

\paragraph{Parallel component}
Let $u\coloneq-\CondDeriv>0$ and $c\coloneq-\CondDerivMean$, so the down-crossing-weighted
marginal is $q_z(u)\propto u\,r(u)^{-3}$, where $r(u)\coloneq\sqrt{(u-c)^2+\alpha_{\mathrm{eff}}^2}$.

\emph{Step 1: find an antiderivative.} Split $u=(u-c)+c$ and integrate each piece
separately. With $s\coloneq u-c$, the first piece is an exact derivative,
$\int s\,(s^2+\alpha_{\mathrm{eff}}^2)^{-3/2}\dif s=-(s^2+\alpha_{\mathrm{eff}}^2)^{-1/2}=-1/r$;
the second is the standard arctangent-free form
$\int(s^2+\alpha_{\mathrm{eff}}^2)^{-3/2}\dif s=s/(\alpha_{\mathrm{eff}}^2 r)$. Together,
\begin{align}
  F(u) \coloneq -\frac{1}{r(u)} + \frac{c(u-c)}{\alpha_{\mathrm{eff}}^2\,r(u)}
      = \frac{c(u-c)-\alpha_{\mathrm{eff}}^2}{\alpha_{\mathrm{eff}}^2\,r(u)}
\end{align}
is an antiderivative of $u\,r(u)^{-3}$, hence proportional to the CDF of $q_z$.

\emph{Step 2: evaluate the endpoints.} At $u=0$, $r(0)=\sqrt{c^2+\alpha_{\mathrm{eff}}^2}=R$, giving
$F(0)=(-c^2-\alpha_{\mathrm{eff}}^2)/(\alpha_{\mathrm{eff}}^2 R)=-R/\alpha_{\mathrm{eff}}^2$. As
$u\to\infty$, $r(u)\sim u$, giving $F(u)\to c/\alpha_{\mathrm{eff}}^2$. So $F$ increases
monotonically (as any CDF must) from $-R/\alpha_{\mathrm{eff}}^2$ to $c/\alpha_{\mathrm{eff}}^2$
as $u$ ranges over $(0,\infty)$.

\emph{Step 3: rescale to a uniform draw.} Define $h\coloneq\alpha_{\mathrm{eff}}^2 F(u)$, so by
Step 2, $h$ ranges over exactly $[-R,\,c]$ as $u$ ranges over $(0,\infty)$---and since
$F$ is (up to the constant rescaling by $\alpha_{\mathrm{eff}}^2$) the CDF of $q_z$, $h$ is
uniformly distributed on $[-R,c]$ when $u\sim q_z$. Drawing $\xi_1\sim\mathrm{Unif}(0,1)$
and setting $h=\xi_1(c+R)-R$ therefore produces a correctly-distributed $h$; the
sampling problem has reduced to inverting Step 1's formula for $u$ given $h$.

\emph{Step 4: invert for $u$.} Writing $q\coloneq u-c$ and squaring
$h\,r(u) = cq-\alpha_{\mathrm{eff}}^2$ gives a quadratic in $q$,
\begin{align}
  (c^2-h^2)\,q^2 - 2c\alpha_{\mathrm{eff}}^2\,q - \alpha_{\mathrm{eff}}^2(h^2-\alpha_{\mathrm{eff}}^2) = 0,
\end{align}
whose discriminant simplifies to $4\alpha_{\mathrm{eff}}^2h^2(R^2-h^2)$ using
$R^2=c^2+\alpha_{\mathrm{eff}}^2$. Of the two roots, the one consistent with $r(u)>0$ is
\begin{align}
  q = \frac{c\,\alpha_{\mathrm{eff}}^2+\alpha_{\mathrm{eff}}\,h\sqrt{R^2-h^2}}{c^2-h^2}.
\end{align}
As a check, at $h=-R$ (the lower endpoint, corresponding to $u=0$) this gives
$q=c\alpha_{\mathrm{eff}}^2/(c^2-R^2)=-c$, i.e.\ $u=0$, as required.
For numerical evaluation, we use the equivalent form
\begin{align}
  q = \alpha_{\mathrm{eff}}\,
  \frac{\alpha_{\mathrm{eff}}\sqrt{R^2-h^2}+c\,h}
       {c\sqrt{R^2-h^2}-\alpha_{\mathrm{eff}}\,h},
\end{align}
which is well defined at $h=-c$ for $c>0$.
This is the form used in Algorithm~\ref{alg:sampleqz_ggx}.
\hfill$\square$

\paragraph{Perpendicular component}
Conditioned on $\CondDeriv$ and precision $\tau$, the perpendicular residual is
$\GradPerp\mid\CondDeriv,\tau \sim \mathcal{N}(\CondMeanPerp,\CondCovPerp/\tau)$, with
$\CondMeanPerp,\CondCovPerp$ the same conditional mean and covariance computed by
ordinary Gaussian conditioning on $\CondDeriv$ already used in
\cref{alg:samplevndf}. Marginalizing $\tau\mid\CondDeriv\sim\mathrm{Gamma}(3/2,\beta)$,
$\beta=1+(\CondDeriv-\CondDerivMean)^2/(2\CondDerivVar)$, gives a residual
$\boldsymbol\eta=\GradPerp-\CondMeanPerp$ following the same power-law form as
\cref{eqn:ggx_grad} one dimension down:
\begin{align}
  p(\boldsymbol\eta)\propto\bigl(1+\boldsymbol\eta^\mathsf{T}\CondCovPerp^{-1}\boldsymbol\eta/(2\beta)\bigr)^{-5/2}.
\end{align}
The linear change of variables $\boldsymbol\eta = \boldsymbol L\boldsymbol\eta'$, with
$\boldsymbol L$ a Cholesky factor of $\CondCovPerp=\boldsymbol L\boldsymbol L^\mathsf{T}$, reduces
this to an isotropic radial distribution, whose CDF inverts by the same substitution
used for \cref{eqn:mixture_projarea} one dimension up, giving us the $\rho,\varphi$
update and the $\boldsymbol L,\GradPerp$ steps in Algorithm~\ref{alg:sampleqz_ggx}.
\hfill$\square$

\end{document}